\documentclass[traditabstract]{aa} 

\usepackage{graphicx}
\usepackage{subcaption}
\usepackage{txfonts}
\usepackage{ulem}
\usepackage{xcolor}
\usepackage{hyperref}

\begin{document}

    \title{Contribution of polycyclic aromatic hydrocarbon ionization to neutral gas heating in galaxies: model versus observations\thanks{The python code for the full model is available at \url{https://gitlab.irap.omp.eu/COSMIC-PAH/photoelectric-heating/tree/master}}}

   \author{O. Bern\'e,
          \inst{1}
          S. Foschino 
          \inst{1}
          \and 
          F. Jalabert
          \inst{1}
          \and
          C. Joblin
          \inst{1}
          }

   \institute{Institut de Recherche en Astrophysique et Planetologie, Universit\'e Toulouse III - Paul Sabatier, CNRS, CNES,\\
   9 Av. du colonel Roche, 31028 Toulouse Cedex 04, France\\
              \email{olivier.berne@irap.omp.eu}
             }

   \date{Received ?? ; accepted ??}

\titlerunning{Gas heating by PAHs} 
 
  \abstract
  {
 The ionization of polycyclic aromatic hydocarbons (PAHs), by ultraviolet (UV) photons from massive stars is expected to account for a large fraction of the heating of neutral gas in galaxies. Evaluation of this proposal, however, has been limited by {our} ability to directly compare observational diagnostics
 to the results of a molecular model describing PAH ionization.
{The objective of this article is to take advantage of the most recent values of molecular parameters derived from  laboratory experiments and quantum chemical calculations on PAHs and provide a detailed comparison between modeled values and observational diagnostics for the PAH charge state and the heating efficiency for PAHs.}
{Despite the use of a simple analytical model, we obtain a good agreement between model results and observational diagnostics} over a wide range of radiation fields and physical conditions, in environments such as star-forming regions, galaxies, and protoplanetary disks. In addition, we found that the {modeled photoelectric heating rates by PAHs} are close to the observed cooling rates given by the gas emission. These results show that PAH ionization 
 is the main source of neutral gas heating in these environments. The results of our {photoelectric heating model by PAHs} can thus be used to assess the contribution of UV radiative heating in galaxies  ({\it vs} e.g. shock). 
 We conclude on the importance of implementing the physics of PAH ionization in astrophysical codes, which are developed, for example, for the evaporating surfaces of protoplanetary disks, the diffuse interstellar medium, and the photodissociation regions associated with star-forming regions in the local and distant universe. We provide the empirical formulas to calculate the heating rates and heating efficiencies for PAHs in this article. 
}

   \keywords{}
   \maketitle
%

\section{Introduction}

In a seminal paper on the temperature of interstellar gas,
it was proposed by \citet{spit48} that the photoelectric 
(PE) effect on dust grains provides an important source of heating of the neutral 
gas in galaxies. 
The central idea behind this PE heating mechanism is that the interaction with dust of ultraviolet (UV) photons from young stars, with energies up to the Lyman limit ($E=13.6$ eV), releases electrons that can carry a couple eV of kinetic energy and thermalize with the 
gas, resulting in gas heating. Following the initial proposal of 
Spitzer, other authors proposed more detailed models of this mechanism 
\citep{jur76,dej77,dra78}. After the discovery that large molecules 
of the family of polycyclic aromatic hydrocarbons (PAHs) are ubiquitous 
in space \citep{leger_puget84,allamandola_polycyclic_1985}, 
it was proposed that these large molecules could contribute to
the heating of the HI gas \citep{dhe87,lep88}.
\citet{Verstraete1990} measured the photoionization yield of the PAH coronene, and estimated the contribution of a PAH population with a typical size of 80 carbons to the PE heating. They concluded that PE effect on PAHs represents a major contribution to the gas heating in cold diffuse clouds and a significant contribution in the warm phase.
\citet{bakesandtielens94} and \citet{weingartner_photoelectric_2001} 
(respectively BT94 and WD01 hereafter) developed models
that take into account grain size distributions {from classical dust sizes (up to typically $\sim$0.1\,$\mu$m)} down to the molecular (PAH) domain. BT94 concluded that about half of the gas heating is due to grains containing less than 1500 C atoms. 
While a proper description of PE heating in dust models is central to a number of astrophysical models and analysis, its implementation has only been benchmarked against observations in a limited number of studies. Deriving values of the PE heating efficiency {for dust}, $\epsilon$, from observations requires quantifying dust emission but also the total neutral gas cooling power from all relevant infrared (IR) cooling lines {in particular from atomic oxygen and ionized carbon.} Several studies have used observations from \textit{Herschel}, SOFIA, and 
\textit{Spitzer} to derive the cooling budget of the neutral gas \citep{bernard-salas_spatial_NGC7023_2015,pab21,Joblin2018}. Recent studies of Galactic photodissociation regions (PDRs) associated with star-forming regions have compared observationally derived $\epsilon$ values with those calculated with the model of BT94 \citep{salgado_orion_map_eps_2016,sal19,pab21}.
There is a good qualitative agreement in terms of the shape of the curves but the calculated values are found to be larger than the observed ones by at least a factor of three. 
Since the earlier proposal that PAHs may provide a major contribution to the PE heating \citep{dhe87,lep88,Verstraete1990}, a number of observations have revealed a tight correlation between the [CII] flux and the PAH emission flux, which is in agreement with this proposal \citep{hel01, joblin_gas_2010, leb12}. \citet{joblin_gas_2010} mentioned that the PAH charge varies over the NGC\,7023 nebula \citep{rapacioli_spectroscopy_2005, berne_analysis_2007} and this is expected to affect the contribution of PAHs to PE heating. Following this idea, \citet{okada_probing_2013} provided 
the first study attempting to assess the connection between PAH charge and PE heating combining \textit{Spitzer} and \textit{Herschel} observations.
The authors derived values of the ratio of gas emission to the sum of gas and PAH emission {in several PDRs}, and showed that its variation with the ionization parameter can be rationalized with a simple prescription based on previous modeling works \citep{Tielens_book05, weingartner_photoelectric_2001}. 
This study concluded that PAHs provide a dominant contribution to the PE heating. However, it was conducted on unresolved sources which provided limited accuracy on the derived values and did not include any modeling of PAH ionization to support the conclusions.
Modeling the charge state of PAHs and their contribution to the PE heating of the gas relies on the knowledge of the molecular properties of these species. Part of the relevant data have been gathered into models that describe the variation of the charge state of PAHs \citep[e.g.][]{montillaud_evolution_2013, Andrews2016}. In addition, key molecular parameters that are needed to model {the PE heating rate by PAHs}  of relevant sizes (i.e. > 30 carbon atoms), both in neutral and cationic forms, are the photo-absorption cross sections $\sigma_{\rm abs}$,  the ionization yields $Y$, and the ionization potentials (IPs). {The earlier values of $Y$ obtained for neutral PAHs \citep{Verstraete1990,Jochims1996} are now completed with values for cations \citep{wenzel2020}.} An additional fundamental parameter is the partition coefficient \citep{dhe87,Verstraete1990},
here written $\langle \gamma_e (E) \rangle$, which determines, upon ionization by a photon of energy $E$, the fraction of the energy that is carried away in kinetic 
energy by the photoelectron relative to the available energy $E-IP$. In this regard, photoelectron spectral images upon UV ionization of coronene have been reported by \citet{brechignac_photoionization_2014}.

In this article we present a study in which we evaluate 
in detail the importance of the ionization of PAHs as a source 
of neutral gas heating in galaxies.
We first derive in Sect.~\ref{sec_observations} the observational
diagnostics on the PAH charge state and {the gas heating efficiency for PAHs}.  
In Sect. ~\ref{sec_model}, we present 
a simple analytical model of PAH ionization and gas heating {by the PE effect on PAHs}, which includes state-of-the-art data for molecular parameters as provided by laboratory measurements and quantum chemical calculations.
In Sect.~\ref{sec_mod_results_and_comp} we present the model results and comparisons to the models of WD01 and BT94.
In Sect.~\ref{sec_model-vs-obs}, we compare the observational diagnostics with results obtained with our PAH model.  
In Sect.~\ref{sec_discussion}, we discuss the dominant role 
of PAH ionization as a source of gas heating, before concluding in Sect.~\ref{sec_conclusion}.

\section{Observational diagnostics of PAH charge state and efficiency of gas heating}
\label{sec_observations}

In order to characterize the contribution of PE
effect on PAHs to the gas heating, we derive two diagnostics from observations.
The first diagnostic we are interested 
in is the ionization fraction of PAHs as derived from observations, 
$R_{\rm i}$, which is determined by:
\begin{equation}
    R_{\rm i}=\frac{I_{\rm{PAH}^+}}{I_{\rm{PAH}^+}+I_{\rm{PAH}^0}},
    \label{eq_r_i}
\end{equation}
where $I_{\rm{PAH}^+}$ and $I_{\rm{PAH}^0}$ are the IR emission intensities attributed to cationic and neutral PAHs, respectively. 
{In this study, we use the
generic tool called PAHTAT} \citep{pilleri_evaporating_2012}
that allows one to derive  $I_{\rm{PAH}^+}$ and $I_{\rm{PAH}^0}$ from 
spectroscopic IR observations\footnote{PAHTAT is available on the Cosmic PAH portal at \url{https://cosmic-pah.irap.omp.eu/}}.

The second diagnostic we derive, following \citet{okada_probing_2013}, 
is the ratio of gas emission to the sum of gas and PAH emission:
\begin{equation}
    R_{\rm e}=\frac{I_{\rm gas}}{I_{\rm PAH}+I_{\rm gas}}, 
\label{eq_r_e}
\end{equation}

with $I_{\rm PAH}$ the intensity of PAH emission and $I_{\rm gas}$ the intensity of neutral gas cooling lines (dominated by the far-infrared [CII] and [OI] lines, see Sect.~\ref{app_derivation_gastopah_NGC7023}). 
When $I_{\rm PAH} >> I_{\rm gas}$, $R_{\rm e} \approx  \frac{I_{\rm gas}}{I_{\rm PAH}}$.

In addition to the {two observational diagnostics described above,} we derive the value of the 
ionization parameter:
\begin{equation}
\gamma=G_0 \times \sqrt{T}/ n_{\rm e},
\end{equation}
where $T$ is the gas temperature in K, and $n_{\rm e}$ is the electron 
density in cm$^{-3}$. $G_0$ quantifies the intensity of the UV radiation field:
\begin{equation}
    G_0=\frac{1}{Z}\int_0^{\Omega} \int_{91.2 \rm{nm}}^{240\rm{nm}}I_{\lambda}(\lambda)~d\lambda~d\Omega,
    \label{eq_G0}
\end{equation}
where $I_{\lambda}(\lambda)$ is the intensity of the radiation field in
the Solar neighborhood as given by \cite{Habing68} (here in W~m$^{-1}$~sr$^{-1}$~$\rm{nm}^{-1}$),
$\Omega$ is the solid angle of this radiation field and $Z=1.68 \times 10^{-6}$\,W~m$^{-2}$ is the normalization factor. This definition of $G_0$ is equivalent to that of \citet{le_petit_model_2006}. The interval of integration over wavelengths of Eq.~\ref{eq_G0} corresponds to an energy range between 13.6 and 5.17\,eV. 
The two observational diagnostics ($R_{\rm i}$ and $R_{\rm e}$) 
and $\gamma$ are derived for the NGC 7023 nebula in spatially resolved observations, and for a sample of Galactic and extragalatic sources described in the following sections.

\subsection{{The spatially resolved NGC 7023 NW PDR}}
\label{subsec_n7023}

\begin{figure}
\centering
  \includegraphics[width=\hsize]{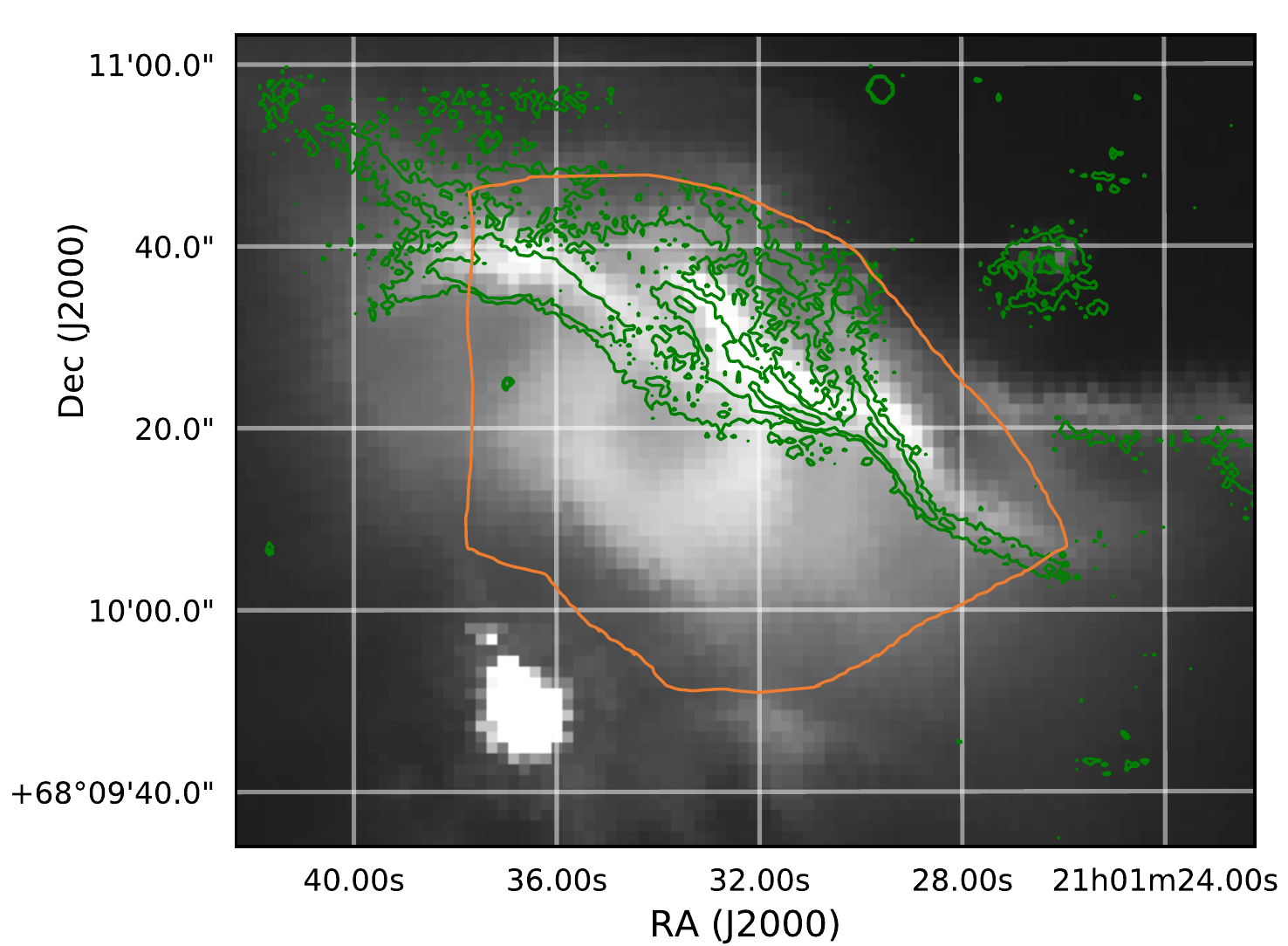}
\caption{General view of the NGC 7023 NW PDR. Grey image is Spitzer-IRAC 
data at 8\,$\mu$m. Green contours show the emission in the H$_2$ 1-0 S(1) line at 0.5, 1 and 2$\times 10^{-7}$\,W\,m$^{-2}$\,sr$^{-1}$, from observations at CFHT presented in \citet{lemaire_high}. 
The orange contour defines the region considered {here for the PE heating} study. }
  \label{fig_NGC7023}

\end{figure}

NGC 7023 is a reflection nebula illuminated by the HD\,200775 double system of spectral types 
B3V and B5 \citep{Finkenzeller85_HD200775,Racine_HD200775} situated at 320\,pc from the Sun 
\citep{Benisty2013}.
{We are interested in the PDR which is located at 42'' north west (NW) of the star.} NGC~7023 NW PDR is the most studied PDR and is seen almost edge-on. The dissociation front (i.e. the interface where H$_2$
is dissociated to 2 H atoms) 
is well seen (Fig.~\ref{fig_NGC7023}) through bright emission in the H$2$ (1-0) $S(1)$ line at 2.12\,$\mu$m,
showing dense filamentary structures ($n_{\rm H}\sim10^{5-6}$cm$^{-3}$) embedded in a more diffuse gas ($n_{\rm H}\sim10^4$cm$^{-3}$) 
\citep{Joblin2018,lemaire_high,Fuente1996_filament,Chokshi88}.
Several estimates point to a UV radiation field intensity at the dissociation front of G$_0$=2600 
\citep{pilleri_evaporating_2012,Joblin2018,Chokshi88}. 
We derive the values of $\gamma$ in NGC 7023 as a function of the distance to the star
(see Appendix~\ref{app_derivation_gamma_NGC7023}) and
summarize these values in Table~\ref{tab_parametresphysiques}. We extract the values of  $R_{\rm i}$ and $R_{\rm e}$
(see Appendix~\ref{app_derivation_fion_obs_ngc7023} and Appendix~\ref{app_derivation_gastopah_NGC7023},
respectively), over the area delimited in Fig.~\ref{fig_NGC7023}.

\begin{figure}[!ht]
    \centering
    \includegraphics[width= 8cm]{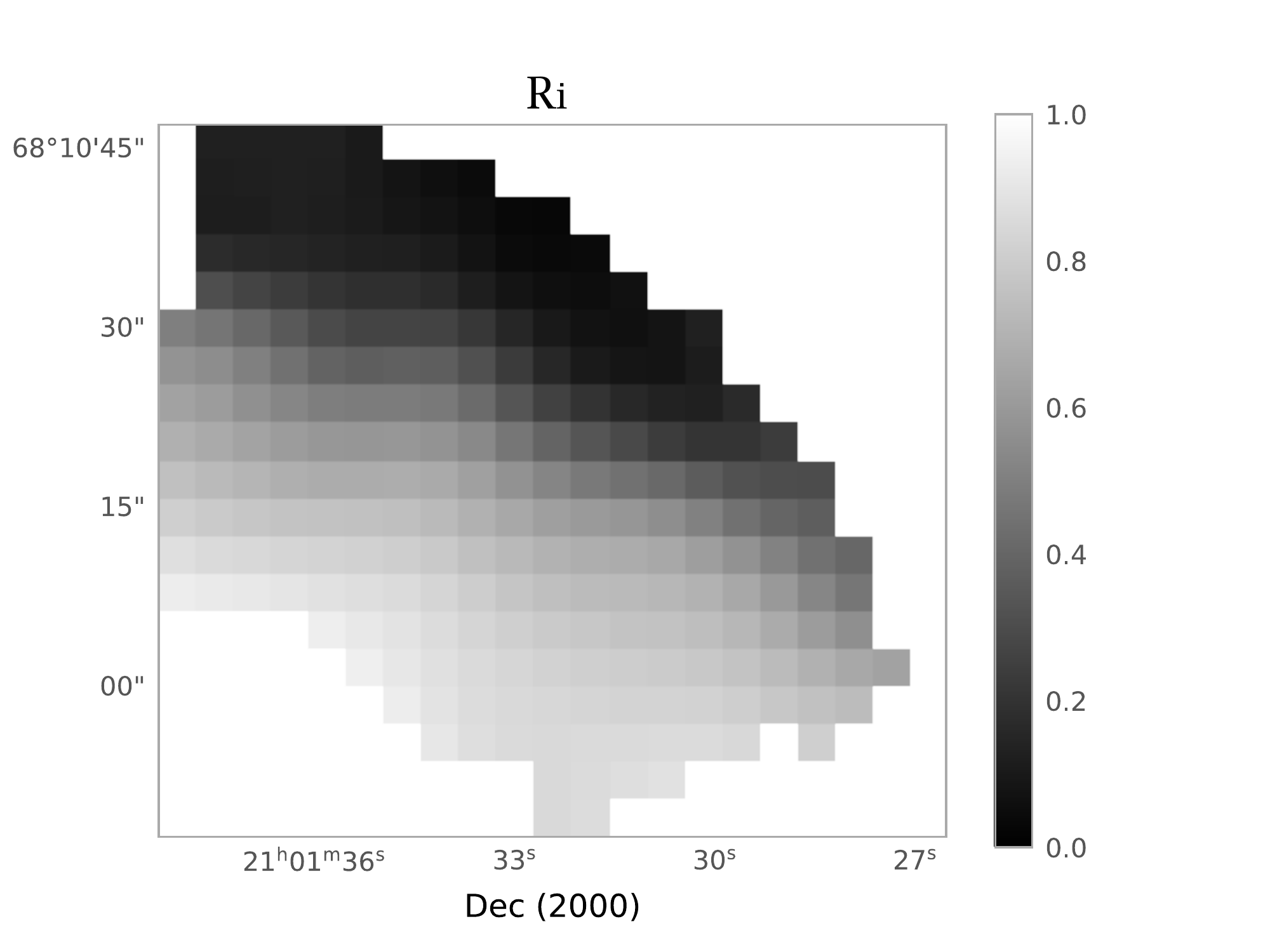}
    \includegraphics[width= 8cm]{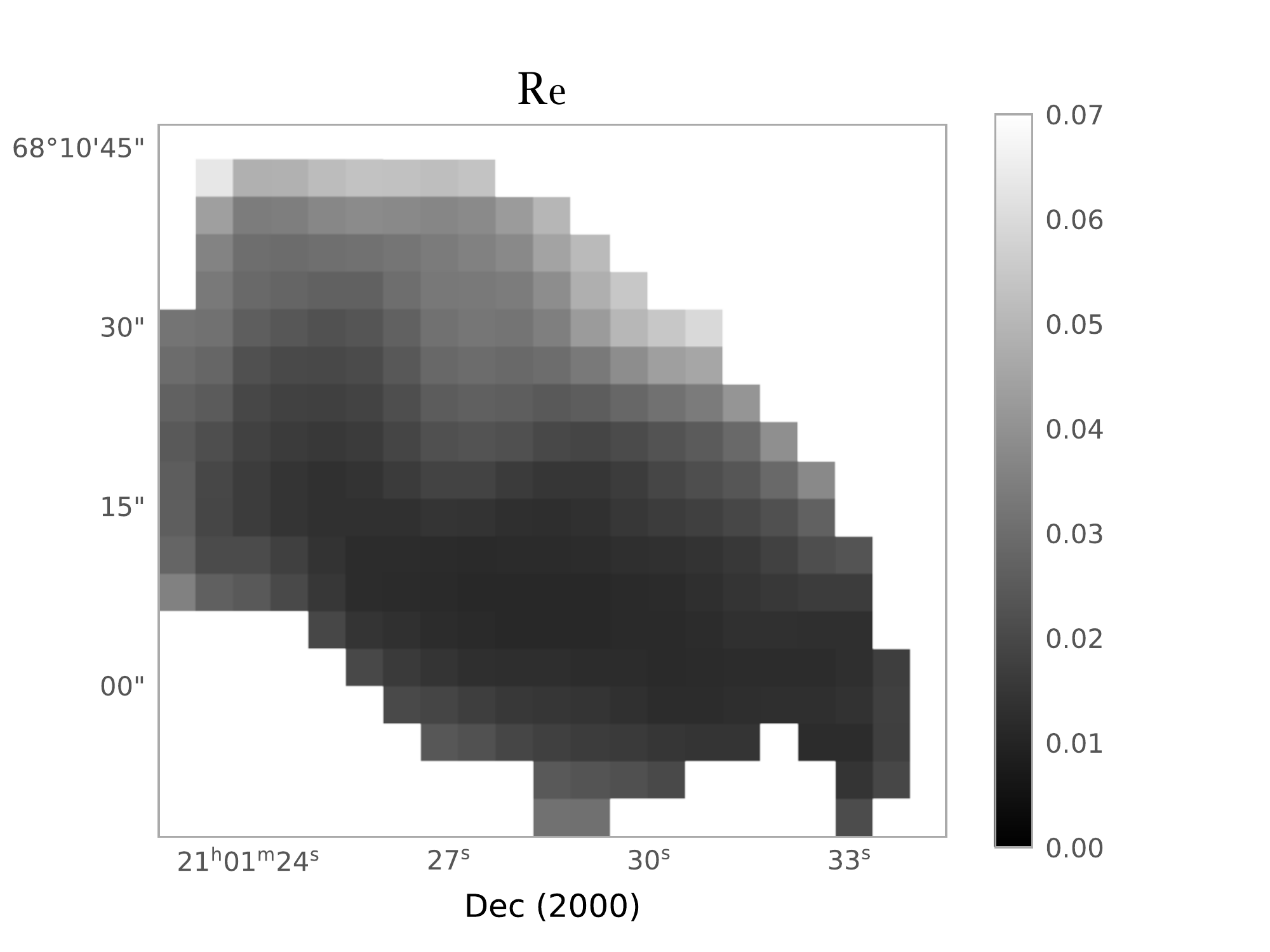}
    \caption { Maps of the ionization fraction of PAHs in NGC 7023 (upper panel), and 
    ratio of gas cooling to PAH emission (lower panel).}
    \label{fig_maps-efficiency-ionization}
\end{figure}

The resulting maps are presented in Fig.~\ref{fig_maps-efficiency-ionization}.
The derived ionization fraction for PAHs is found to decrease from regions near HD\,200775 to regions beyond the dissociation front as the radiation field intensity decreases, as expected.
This result is consistent with earlier studies \citep[e.g.,][]{pilleri_evaporating_2012,boersma_charge_2016}. 
$R_{\rm e}$ follows the opposite trend, that is it increases with 
distance to the star. This is {in line with the trend observed} by \citet{okada_probing_2013}
for three position in NGC 7023. Since PAH emission is directly related to the intensity of the radiation field, this variation of $R_{\rm e}$ is indicative of a better coupling between the radiation field and the gas heating in regions 
far from the star, where $\gamma$ is low and PAHs are mainly neutral. 

\begin{table*}[ht!]
    \centering
        \caption{Adopted physical conditions, and derived values of the ionization parameter $\gamma$ 
    for eight angular distances from the star HD 200775 in NGC~7023 NW.}
    \begin{tabular}{|c|c c c c c c c c |}
    \hline
        d ('') & 20&25&30&35&40& 45& 50&55\\
            \hline
        $G_0$ & 11004&7042&4890&3593&2751&2173&1760&733 \\
    \hline
        T (K) & 750&750&750&750&750&750&350&150\\
    \hline
        $n_{\rm H}$ ($10^4$cm$^{-3}$)& 1.5&1.5&1.5&1.5&2&2&2&2\\
    \hline
        $\gamma$\,(K$^{1/2}$cm$^3)$ & 125567&80363&55807&41001&23543&18602&10293&2807\\ 
    \hline
        
    \end{tabular}
    \label{tab_parametresphysiques}
\end{table*}

\subsection{Sample of Galactic and extragalactic sources}
\label{subsec_source-sample}

\begin{figure}[!ht]
    \centering
     \includegraphics[width=\hsize]{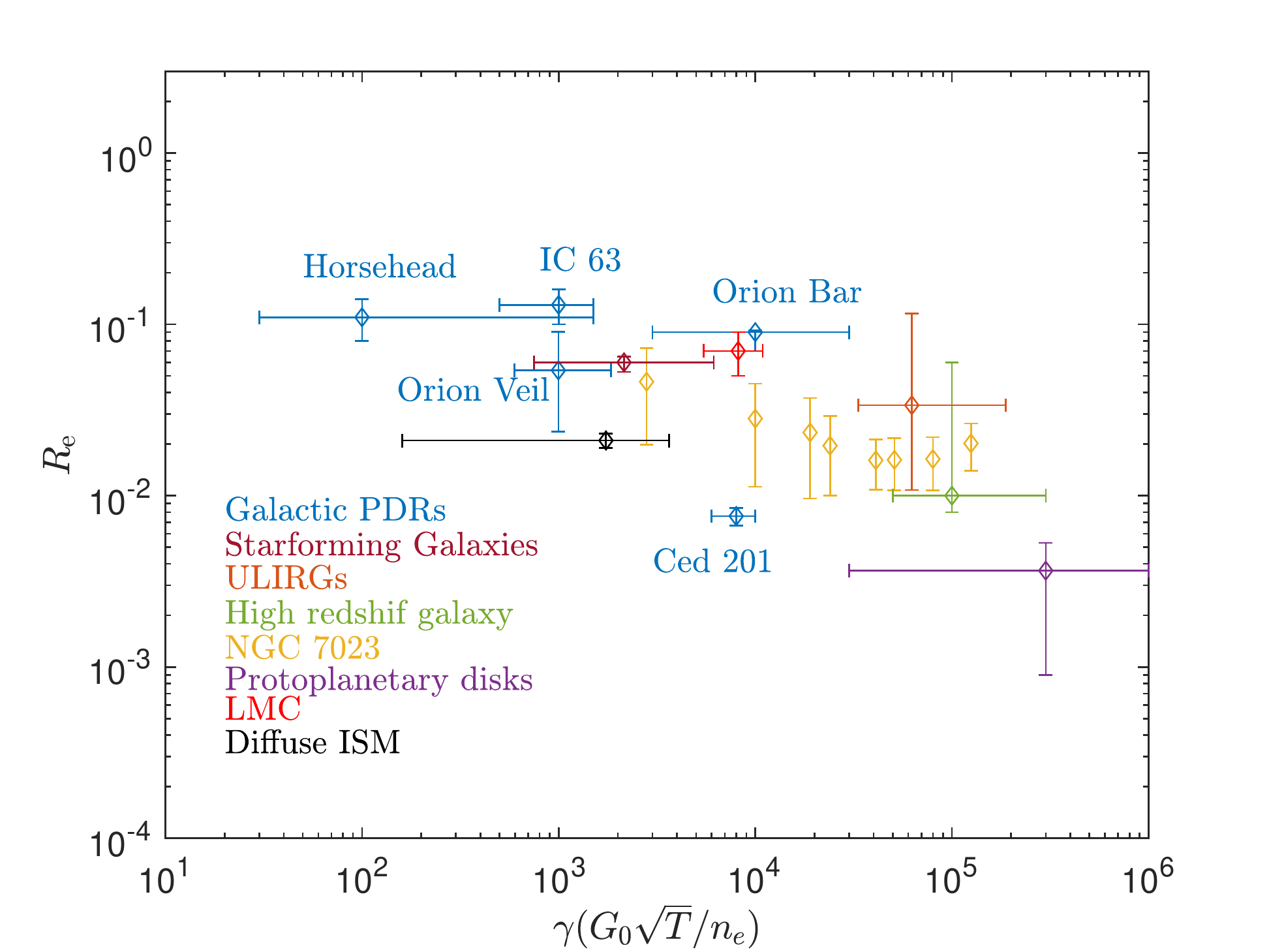}
    \caption{Measured emission ratio $R_{\rm e}$ (Eq.~\ref{eq_r_e}), as a function of physical 
    conditions traced by the ionization parameter $\gamma$, in a sample of astrophysical objects.}
    \label{fig_All-objects_obs}
\end{figure}

Values for $R_{\rm e}$ are derived from observations for a number of astrophysical environments for which we expect the PE effect on PAHs to be a major contributor to gas heating.
These include star-forming regions, protoplanetary disks, star-forming galaxies, ultra luminous infrared galaxies, and a high-redshift galaxy. 
The methodology to derive these parameters is detailed in 
Appendix~\ref{app_derivation_gamma_all_objects}. 
All derived values are summarized in Table~\ref{tab_all-objects} {together with associated $\gamma$ values, and  are shown} in Fig.~\ref{fig_All-objects_obs}.
Similar to the trend observed in NGC\,7023~NW, $R_{\rm e}$ {values are} found to be lower in sources with high $\gamma$ values, such as protoplanetary disks or galaxies that actively form stars. However, the trend is also highly variable. For example,  the Orion Bar and Ced 201 have similar $\gamma$ values but their values of $R_{\rm e}$ differ by an order of magnitude.

\begin{table*}[]
\centering
\caption{Summary of values of $R_{\rm e}$  and ionization parameter 
$\gamma$ derived from observations. \label{tab_all-objects}}
\begin{tabular}{ccc} 
\hline 
Object  &  $R_{\rm e}$  & $\gamma$ \\
\hline 
\multicolumn{3}{c}{Galactic photodissociation regions}\\
\hline
Orion Veil & 0.054 (0.024,0.091) &  997 (595,1850) \\
Orion Bar & [0.07-0.11] & $[0.7\times 10^4 - 2 \times 10^4]$\\
IC 63 & 0.14 (0.11,0.17) & [500-1500] \\
Ced 201 & 0.018 (0.012,0.024) & [6000 - 10 000]\\
Horsehead & 0.11 (0.08,0.14) & [30-1400] \\
\hline 
\multicolumn{3}{c}{Other environments}\\
\hline 
Disks around HAeBe stars & 0.0037 ($9.0\times10^{-4},0.0053$)&  $[3\times 10^4 - 3 \times 10^6]$ \\
Diffuse ISM & $0.021 \pm 0.002$ & [1580 - 1900]\\
LMC & [0.05-0.1] & [5460 - 10 900] \\
Star-forming Galaxies & [0.0048 - 0.0073] & $[1.4\times 10^3 - 4 \times 10^3]$ \\
ULIRGs & 0.034 (0.011, 0.12)& $6.3\times 10^{4} ({3.4 \times 10^{4}, {1.9 \times 10^{5}}})$\\
High $z$ Galaxy & [0.008-0.022] & $[5\times 10^4 - 2 \times 10^5]$ \\
\hline
\end{tabular}
\tablefoot{{The listed values are either median with confidence intervals (see Appendix~\ref{app_derivation_gamma_all_objects} details on the boundaries of these intervals), or provide the range 
of considered values in brackets.}}
\end{table*}

\section{{Molecular model of the charge state and photoelectric emission from PAHs}}\label{sec_model}

Here we present a simple molecular model, whose objective is to compute the PAH ionization fraction, {the heating efficiency for PAHs, and the heating rates by PAHs.} This model is based on the photophysics of isolated interstellar PAHs and includes the latest results
from laboratory experiments for PAH molecular parameters. 
Upon absorption of a UV photon, three main processes are expected to be in competition to relax the energy: 
ionization, dissociation, and radiative cooling {\citep[see, e.g.,][]{Joblin2020}}. 
Photoelectric heating is related to ionization, and therefore we focus on this process in the following.

\subsection{{Selection of species and molecular parameters}}\label{subsec_molecular-parameters}

{A first step consists in selecting species for which molecular data is available.} 
Here we rely on {information given by} the 
\textit{Theoretical spectral database of PAHs}\footnote{ \url{http://astrochemistry.oa-cagliari.inaf.it/database/pahs.html}} 
\citep{malloci_-line_2007}. The chosen molecules are listed in Table~\ref{tab_listemolecules}, having compact shapes 
and different characteristic sizes {relevant for interstellar} PAHs \citep{leger_puget84,Leger1989_PAHmodel,allamandola_polycyclic_1985,montillaud_evolution_2013, Andrews2016}. 

\begin{table}[ht!]
    \centering
    \begin{tabular}{cc} 
    \hline
    Name & Chemical formula \\
    \hline

       Ovalene & C$_{32}$H$_{14}$ \\ 
       Tetrabenzocoronene  & C$_{36}$H$_{16}$\\
       Circumbiphenyl & C$_{38}$H$_{16}$ \\
       Circumanthracene & C$_{40}$H$_{16}$ \\
       Circumpyrene & C$_{42}$H$_{16}$ \\
        Hexabenzocoronene & C$_{42}$H$_{18}$ \\
       Dicoronylene & C$_{48}$H$_{20}$ \\
       Circumcoronene & C$_{54}$H$_{18}$\\
        Circumovalene & C$_{66}$H$_{20}$ \\
\hline
    \end{tabular}
    \caption{PAH species considered in the present study.}
    \label{tab_listemolecules}
\end{table}

In the present model, we consider four charge states that is $Z \in \{-1,0,1,2\}$). 
Although there is no clear spectroscopic evidence for the presence of anions ($Z\leq-1$) in the ISM, we include them
because they could play a role in gas heating in environments with low $\gamma$ values.  
Cations ($Z=1$) and neutrals ($Z=0$) have long been proposed to exist in PDRs \citep[see review in][]{tielens_interstellar_2008}. 
Theoretical \citep{malloci_-line_2007} and experimental \citep{wenzel2020} data have shown that the dication stage
($Z=2$) is accessible  in PDRs for the considered species (Table~\ref{tab_listemolecules}), because their ionization potential IP$^{2}$ is lower than the Lyman limit of 13.6\,eV. Data for IP$^{3}$ values for PAHs are scarce, but \citet{zhen_vuv_2016} have shown that they are greater than 13.6\,eV for small PAHs, and \citet{wenzel2020} 
demonstrate that this is probably the case up to a typical size of $N_{\rm C} \sim 50$. Therefore, we do not consider trications in our model. 

\begin{figure}[!ht]
    \centering
    \includegraphics[width=\hsize]{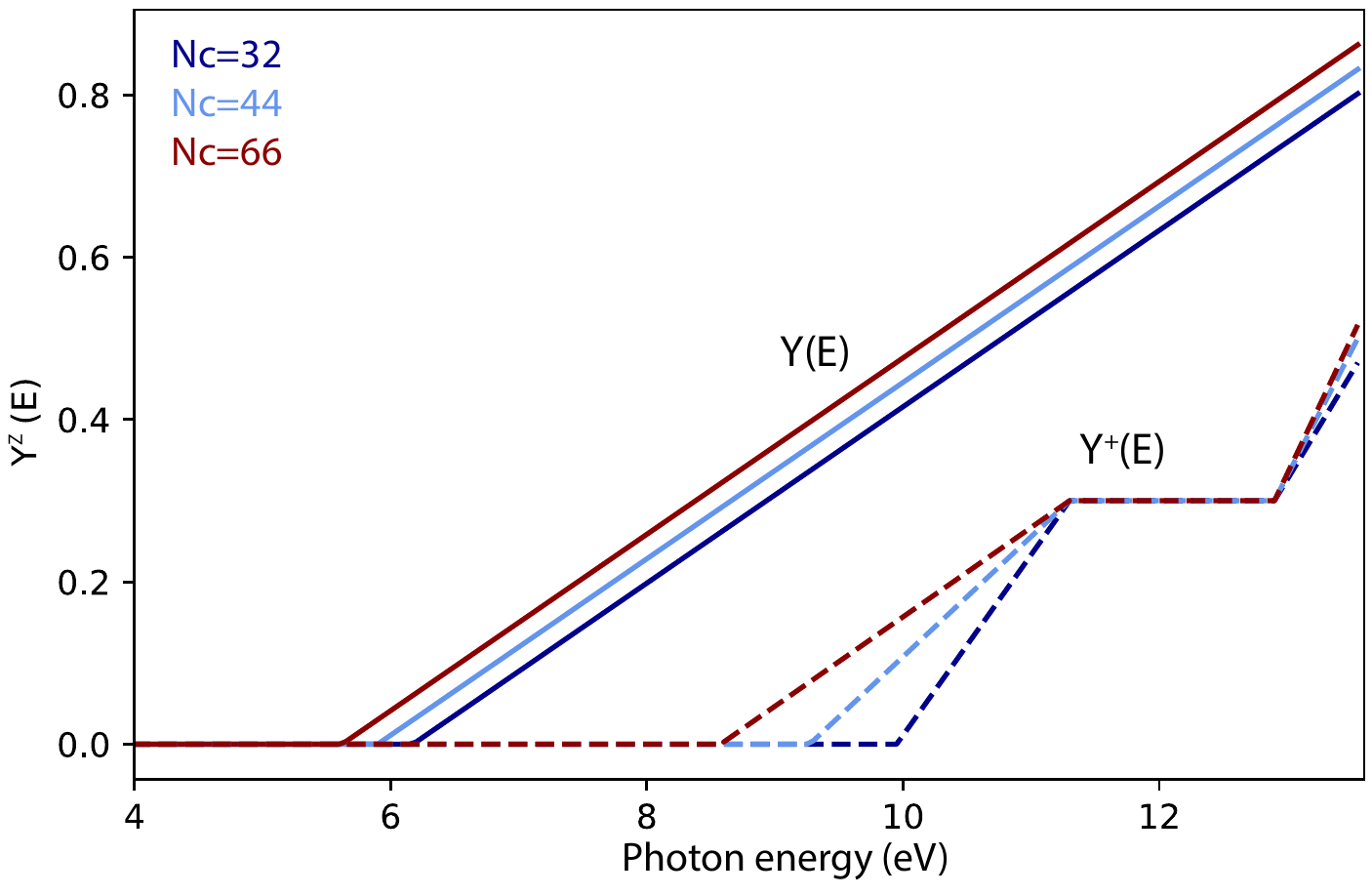}
    \caption{Ionization yields {$Y^{Z}(E)$} for charge states $Z=0$ (solid lines) and $Z=1$ (dashed lines), computed from Eqs.~\ref{eq_yieldjoachims} {and~\ref{eq_yieldwenzel}-\ref{eq_yieldwenzel2}} respectively, and for the three different PAH sizes.  The yield for $Z=-1$, $Y^{-}(E)=1$ is not shown, for clarity of the figure.}   
    \label{fig_yield}
\end{figure}

Values of the ionization potentials {of a PAH with a given size $N_{\rm C}$}, are obtained  using the empirical formalism of WD01, with updated parameters from \citet{wenzel2020} {as given in Appendix~\ref{app_IPs}}.
Several studies provide measurements of the ionization yields of neutral
\citep{Jochims1996,Verstraete1990} and cationic \citep{zhen_vuv_2016,wenzel2020} PAHs. From these measurements,
\citet{Jochims1996} and \citet{wenzel2020} derived empirical laws for the ionization yields of neutral and cationic PAHs, respectively, which are used in this work (see Appendix~\ref{app_IYields}).
The ionization yield curves are shown as a function of incident photon energy in Fig.~\ref{fig_yield} for three PAH sizes. 
{Similarly to \citet{vis07}, we use $Y^{-}(E)=1$. The threshold in energy for the detachment of the electron on anions is given by the electron affinity \citep[e.g.,][]{Tschur06}. The latter is very variable with the molecule but remains below 2 eV \citep[see \textit{Theoretical spectral database of PAHs};][]{Malloci2005, Wahab22}.  We therefore assumed, as an approximation, that $Y^{-}(E)=1$ at all energies. We present another case in Appendix~\ref{app_threshold}.}

\begin{figure}[!ht]
    \centering
    \includegraphics[width=7cm]{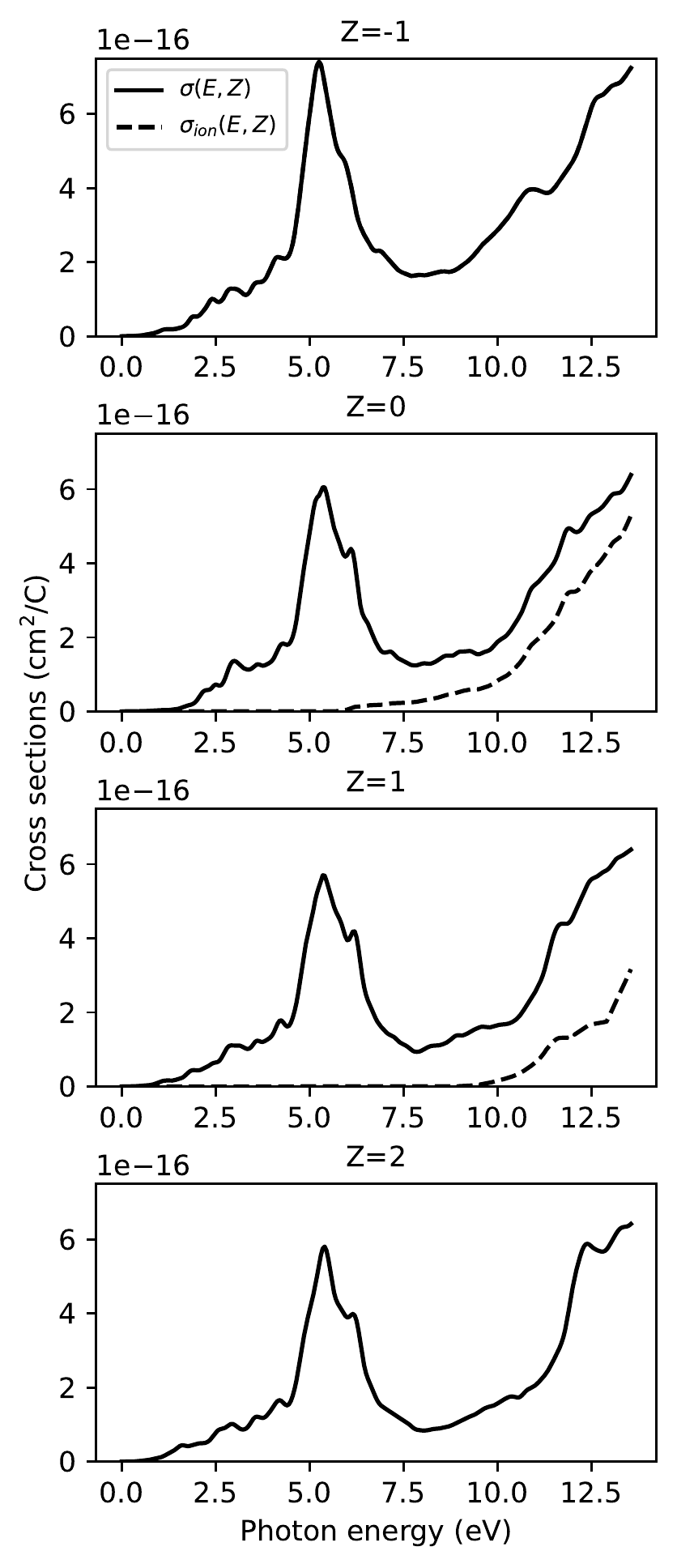}
    \caption{Average photoabsorption ($\sigma$, solid line) and photoionization ($\sigma_{\text{ion}}$, 
    dashed line) cross sections per C atom of PAHs adopted in the 4 energy level model.}   
    \label{fig_sectionsefficaces}
\end{figure}

Finally, we use the photoabsorption cross sections given by the \textit{Theoretical spectral database of PAHs}\footnote{\url{https://www.dsf.unica.it/~gmalloci/pahs/}} \citep[see also][]{malloci04}. 
First, we {retrieve} the photoabsorption cross section for each molecule in Table~\ref{tab_listemolecules} and for each value 
of $Z \in \{-1,0,1,2\}$  and normalize it to the number of carbon atoms $N_{\rm C}$. 
Second, for each value of $Z \in \{-1,0,1,2\}$ we compute the averaged normalized photoabsorption cross section. 
This yields an average photoabsorption cross section per carbon atom $\sigma(E,Z)$, from which the ionization cross section per carbon atom can be derived 
using: 
\begin{equation}
    \sigma_{\text{ion}}(E,Z)=Y^{Z}(E)\times\sigma(E,Z) \text{~~~(cm$^2$ C$^{-1}$)}.
    \label{eg_sigion}
\end{equation}

\noindent Figure~\ref{fig_sectionsefficaces} represents the $\sigma(E,Z)$ and $\sigma_{\text{ion}}(E,Z)$ 
curves as used in the model.

\subsection{PAH charge states and ionization fraction}\label{sec_popfrac}

At equilibrium the fractions $f^{Z}$ of each PAH population in a $Z$ charge state is given by : 
\begin{equation}
    \begin{cases}
        \text{charge level Z = -1 :} &f^-k_\text{det} = f^0 k_\text{att} n_e \\
        
        \text{charge level Z = 0 :} &f^0k_\text{pe}^0 + f^0 k_\text{att} n_e = f^+k_\text{rec}^+ n_e + f^-k_\text{det} \\
        
        \text{charge level Z = 1 :} &f^+k_\text{rec}^+ n_e + f^+k_\text{pe}^+ = f^{2+}k_\text{rec}^{2+} n_e + f^0k_\text{pe}^+ \\
        
        \text{charge level Z = 2 :} &f^{2+}k_\text{rec}^{2+} n_e = f^+k_\text{pe}^+
    \label{eq_chargeequilibre}
    \end{cases}
\end{equation}
where k$_{\text{pe}}^{Z}$ and k$_{\text{rec}}^{Z}$ are respectively the ionization and recombination 
rates of PAHs in the $Z \in\{0,1,2\}$ charge state. We note that when $Z$ is an exponent in the notation, 
it is written $\{-,0,+,2+\}$ rather than $\{-1,0,1, 2\}$, to avoid confusion. 
We also introduce the attachment rate of an electron to a neutral PAH, k$_\text{att}\equiv k_\text{rec}^0$, and the 
detachment rate of the electron from an anion, k$_\text{det}\equiv k_\text{pe}^-$. In addition, {the values of $f^{Z}$ should fulfill the condition:} 
\begin{equation}
    \sum_{Z_{\text{min}}}^{Z_{\text{max}}}f^{Z}=1.
    \label{fraction_normalization}
\end{equation}

\noindent For $Z \in\{-1,0,1,2\}$, Eq. ~\ref{eq_chargeequilibre} and Eq.~\ref{fraction_normalization} yield

\begin{equation}
    f^- = \left(1+\frac{k_\text{det}}{k_\text{att}n_e}+\frac{k_\text{det}k_\text{pe}^0}{k_\text{att}k_\text{rec}^+n_e^2}+\frac{k_\text{det}k_\text{pe}^0k_\text{pe}^+}{k_\text{att}k_\text{rec}^+k_\text{rec}^{2+}n_e^3}\right)^{-1},
    \label{eq_f-}
\end{equation}
\begin{equation}
    f^{0}=\left(1+\frac{k_\text{att}n_e}{k_\text{det}}+\frac{k_\text{pe}^0}{k_\text{rec}^+n_e}+\frac{k_\text{pe}^0k_\text{pe}^+}{k_\text{rec}^+k_\text{rec}^{2+}n_e^2}\right)^{-1},
    \label{eq_f0}
\end{equation}
\begin{equation}
    f^{+}=\left(1+\frac{k_{\rm rec}^+n_e}{k_{\rm pe}^0} + \frac{k_\text{pe}^+}{k_\text{rec}^{2+}}+\frac{k_\text{att}k_\text{rec}^+n_e^2}{k_\text{det}k_\text{pe}^0}\right)^{-1},
    \label{eq_f+}
\end{equation}
\begin{equation}
    f^{2+}=\left(1+\frac{k_\text{rec}^{2+}n_e}{k_\text{pe}^+}+\frac{k_\text{rec}^+k_\text{rec}^{2+}n_e^2}{k_\text{pe}^0k_\text{pe}^+}+\frac{k_\text{att}k_\text{rec}^+k_\text{rec}^{2+}n_e^3}{k_\text{det}k_\text{pe}^+k_\text{pe}^+}\right)^{-1}
    \label{eq_f++}.
\end{equation}

\noindent The ionization rates are given by:
\begin{equation}
    k_{\text{pe}}^{Z}=\int_0^{\Omega} \int_{IP^{(Z+1)}}^{13.6} \frac{\sigma_{ion}(E,Z)I_{\rm E}(E)}{E}dE~d\Omega \text{~~~(s$^{-1}$)},
    \label{eq_ionization_rate}
\end{equation}
where $I_{\rm E}(E)$, is the local intensity of the UV radiation field in W\,m$^2$\,eV$^{-1}$\,sr$^{-1}$.

The values of $\sigma_{ion}(E,Z)$ have been defined above (Eq.~\ref{eg_sigion}, in particular).
{On the other hand,} the recombination rate is calculated using Spitzer's 
formalism \citep{spitzer_physical_2004}, adapted for PAH cations by \citet{Verstraete1990}: 

\begin{equation}
       k_{\rm rec} = 1.28\times10^{-10}N_{\rm C}\sqrt{T}(1+\Phi) \text{~~~(cm$^3$\,s$^{-1}$)},
\label{eq_recombinaisonrate}
\end{equation}
where $T$ is the gas temperature, and 
\begin{equation}
\Phi= \frac{eU}{k_BT} = \frac{1.85\times10^5}{T\sqrt{N_{\rm C}}} \text{~~~(dimensionless)}.
\end{equation}
where e is the electron charge, U is the mean electrostatic potential evaluated at the radius of the PAH considered, defined by $U = \frac{1}{4\pi\epsilon_0}\frac{e}{0.9\times10^{-10}\sqrt{N_{\rm C}}}\text{~~~(J C$^{-1}$)}$,
k$_\text{B}$ the Boltzmann constant and T the gas temperature.

\noindent Eq.~\ref{eq_recombinaisonrate} can be extended to include all charge states $Z>0$ 
in the recombination rate: 
\begin{equation}
       k_{\rm rec}^{Z} = 1.28\times10^{-10}N_{\rm C}\sqrt{T}(1+\Phi\times (1+Z)) \text{~~~(cm$^3$\,s$^{-1}$)}.
\label{eq_recombinaisonrate_code}
\end{equation}

\noindent For k$_\text{att}$, we use the work of \citet{car13}
who provide an empirical fit to their quantum chemistry calculations that is
\begin{equation}
    k_\text{att} = a \left(\frac{T}{300\text{K}}\right)^b {\rm exp}\left(-\frac{c}{T}\right) \text{~~~(cm$^3$\,s$^{-1}$)}.
    \label{eq_attachmentrate}
\end{equation}

\noindent  For the parameters $a, b, c$ we use the values determined by \citet{car13} for coronene, C$_{24}$H$_{12}$ which is 
the largest molecule in this study, that is $a=2.74 \times 10^{-9}$ cm$^3$s$^{-1}$, $b=0.11$, and $c=-1.11$. 
We note however that for all the molecules studied by \citet{car13}, the attachment rate is of 
the order of $\sim 10^{-9}$ cm$^{3}$s$^{-1}$ with only a small dependence on temperature. Hence
this choice is unlikely to affect the model results.
\\
Finally, we write the ensemble of charge fractions:
\begin{equation}
\mathbf{F}=\{f^{++},f^{+},f^{0}, f^{-}\},
\label{eq_ionization_fraction_ensemble}
\end{equation}
and the modeled PAH positive ionization fraction is defined as:
\begin{equation}
f_{\rm i}=f^++f^{2+}.
\label{Eq_fion}
\end{equation}

\subsection{{Photoelectic heating efficiency for PAHs}}
\label{sec_PE_efficiency}

The power injected into the gas via the photoelectrons, $P_{\rm e}$, is given by:
\begin{equation}
        P_{\rm e}= f^-\times P_{\rm e}^- + f^0\times P_{\rm e}^{0} + f^+\times P_{\rm e}^{+} ~~~\text{(W)},
    \label{eq_Pgas}
\end{equation}
where $f^-$, $f^0$ and $f^+$ are the fractions of anionic, neutral and cationic PAHs, {derived from Eqs~\ref{eq_f-}, \ref{eq_f0} and ~\ref{eq_f+}, respectively.} 
$P_{\rm e}^{-}$, $P_{\rm e}^{0}$ and $ P_{\rm e}^{+}$ are the powers injected in the gas by 
photoelectrons ejected from anionic, neutral and cationic PAHs, respectively,
which are defined by:
\begin{equation}
    \begin{aligned}
         P_{\rm e}^{Z}=    & \int_{0}^{\Omega}\int_{IP^{Z+1}}^{13.6}\gamma_{\rm e}(E)(E-IP^{Z+1}) \\
                            & \times \sigma_{\text{ion}}(E,Z)\frac{I_{\rm E}(E)}{E}dE~d\Omega~~~\text{(W)}, \\
    \label{eq_P_PE}
    \end{aligned}
\end{equation}

\noindent where $\gamma_{\rm e} (E)$ is the fraction of the energy that,  after ionization, goes into kinetic energy of the photoelectron.
 
Energy conservation upon ionization requires:
\begin{equation}
E - IP = E^*(E) + E_K(E) + E_K^M(E),
\end{equation}
where {$E^*$ is the internal energy of the PAH} and $E_K^M$ is its kinetic energy, which, due to the much higher mass of the PAH compared to the electron and due to conservation of momentum, is negligible {compared to the kinetic energy of the electron, $E_K$}. 
{Next,} following \citet{dhe87}, we define the
partition coefficient $\langle \gamma_e (E) \rangle$ as the energy averaged $\gamma_e (E)$:
\begin{equation}
\langle \gamma_e (E) \rangle  = \langle \frac{E_K(E)}{E - IP} \rangle.
\end{equation}
{We estimate the value of $\langle \gamma_e (E) \rangle$ from photoelectron spectroscopy 
measurements performed using VUV synchrotron radiation and for the coronene molecule \citep{brechignac_photoionization_2014}. The derived value of $\langle \gamma_e (E) \rangle = 0.46 \pm 0.06~$ (68$\%$ confidence interval) is found to be in agreement with the value of $\langle \gamma_e (E) \rangle=0.5$, {which was used in a number of articles}
\citep[e.g.,][]{Verstraete1990,Tielens_book05,bakesandtielens94} and derived from older measurements on benzene by \citet{ter64} (see Appendix~\ref{app_gamma}).} For anionic PAHs, {we assume that all energy available is taken by the electron. For $IP^0 = 0$, this means that all the energy of the photon is transformed into kinetic energy of the electron.}

{On the other hand,} the power absorbed, mostly (but not only) in the UV, by each PAH charge state, $P_{\rm Rad}^{Z}$, is given by:
\begin{equation}
    P_{\rm Rad}^{Z}=\int_{0}^{\Omega}\int_{0}^{13.6\text{\,eV}}\sigma(E,Z)I_{\rm E}(E)~dE~d\Omega~~~\text{(W)}.
    \label{eq_PUV}
\end{equation}   
The {total} radiative power absorbed by PAHs, $P_{\rm Rad}$, is then defined by:
\begin{equation}
        P_{\rm Rad}= f^-\times P_{\rm Rad}^- + f^0\times P_{\rm Rad}^{0} + f^+\times P_{\rm Rad}^{+}+ f^{2+}\times P_{\rm Rad}^{2+} \text{~~~(W)}.
    \label{eq_Ptot}
\end{equation}

We define the heating efficiency for PAHs, $\epsilon_{\rm PAH}$, as the ratio between the power injected in the gas 
via photoelectrons from PAHs to the total power of the radiation absorbed by PAHs, that is
\begin{equation}
    \epsilon_{\rm PAH}=\frac{P_{\rm e}}{P_{\rm Rad}}.
    \label{eq_efficacite}
\end{equation}

\noindent We note that the derived values for the {radiative power} and heating efficiency depend on $N_{\rm C}$, {through absorption and ionization cross sections, and through recombination rates with electrons.}
For the sake of simplicity this variable has, however, 
not been included in the notation.
{Also, our} definition of the PE heating efficiency for PAHs is similar to {that used by WD01 to compute} their heating efficiency for grains, $\epsilon_{\Gamma}$, but {differs from the definition of
$\epsilon$} by BT94, who {considered only absorbed photons in the 6-13.6\,eV range.} 
{Finally}, the PE heating rate {by PAHs} can be derived using the following relation:
\begin{equation}
    \Gamma_{\rm PAH} = P_e \times \frac{f_{\rm C} [C]}{ N_{\rm C} } = \epsilon_{PAH} P_{\rm Rad} \frac{f_{\rm C} [C]}{ N_{\rm C} } ~~~\text{(W\,H$^{-1}$)},
    \label{eq_heat_rate_model}
\end{equation}
where $f_{\rm C} $ is the fraction of cosmic carbon locked in PAHs, and $[C]$ is the cosmic abundance of carbon relative to hydrogen.
The python code to compute $\Gamma_{\rm PAH}$ and $\epsilon_{\rm PAH}$ with the model described in this section is provided through the Cosmic PAH portal at \url{https://cosmic-pah.irap.omp.eu/} or directly at 
\url{https://gitlab.irap.omp.eu/COSMIC-PAH/photoelectric-heating/tree/master}.

\section{Model results and comparison with previous models}
\label{sec_mod_results_and_comp}

\subsection{Charge balance}

\begin{figure}[!h]
    \centering
    \includegraphics[width=9.0cm]{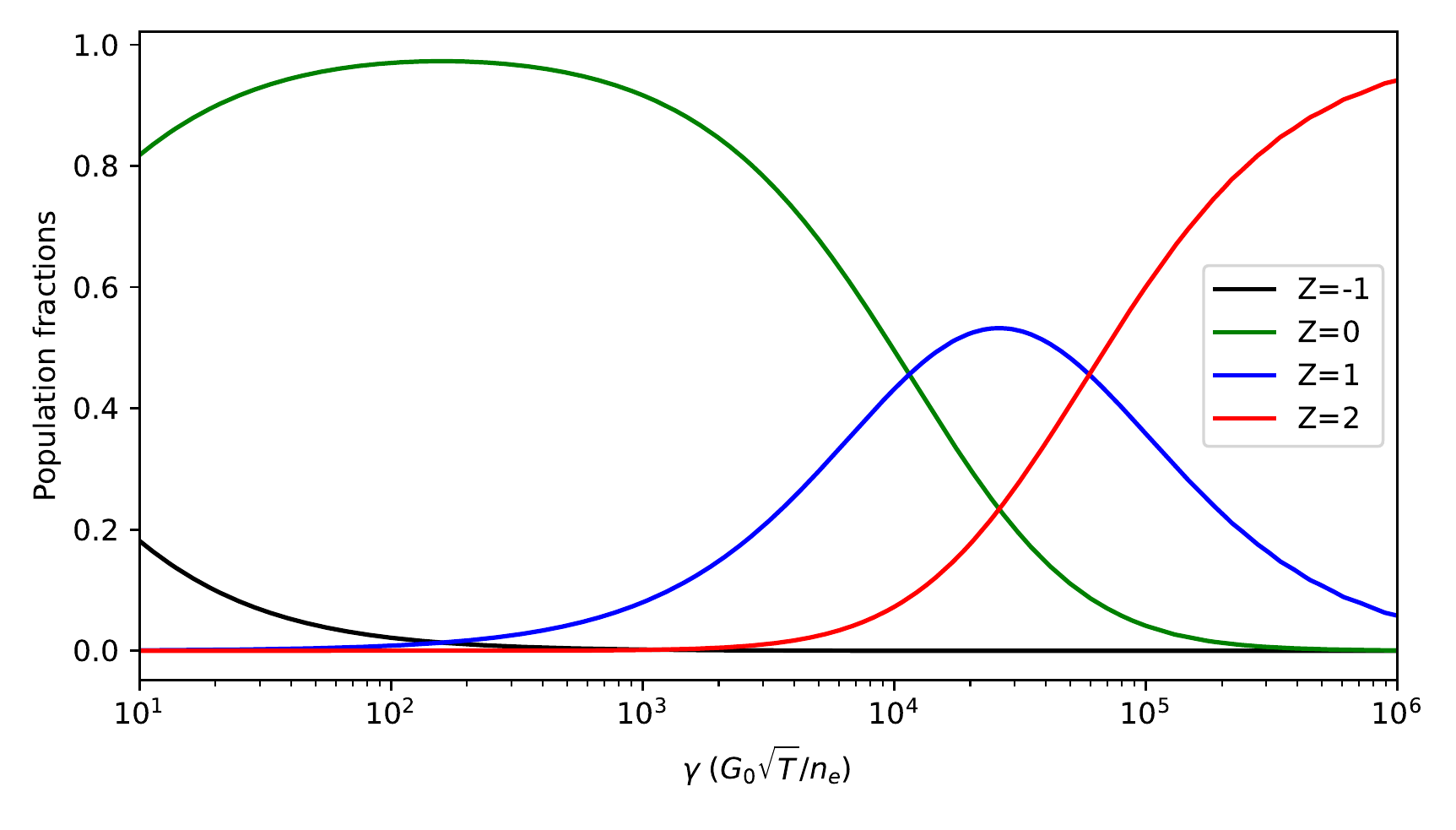}
    \caption{Fractions in the different PAH charge state $Z={-1,0,1,2}$ as a function of the ionization parameter $\gamma$, which were derived using our PAH model and $T_{\rm eff} = 3\times10^4$K.}
    \label{fig_pop_fraction}
\end{figure}

Figure~\ref{fig_pop_fraction} shows the variation of the population fractions of 
PAHs in charge state $Z$, as a function of $\gamma$, derived from the model,
for $T=1000$\,K, $N_{C}=54$, and stellar effective temperature of $T_{\rm eff}=3\times10^4$\,K
using a \citet{Kurucz1993} star model. 
For $\gamma \lesssim 10^3$ (e.g. diffuse ISM or dark clouds), 
the population is dominated by neutral PAHs, with an anionic contribution up to 20\%. A comparable fraction of neutrals and anions was also derived at low $\gamma$ values by both BT94 and WD01 for $5$\,\AA~grains (comparable to PAHs of size $N_C =54$).
Neutral PAHs dominate up to $\gamma \sim 10^4$, followed by cations with a maximum at about $\gamma \sim 3\times 10^4$.
Finally, dications are the dominant species for $\gamma \gtrsim 10^5$. 
For $\gamma=10^5$ and the same stellar effective temperature, WD01 find 
values for the charge fractions (Eq.~\ref{eq_ionization_fraction_ensemble})
of $\mathbf{F}=\{0.7, 0.2, 0.1, 0.0\}$.
We infer comparable fractions but with less dications and more cations,
$\mathbf{F}=\{0.5, 0.4, 0.1, 0.0\}$ (Fig.~\ref{fig_pop_fraction}).
The difference results from the higher ratio of ionization rate to recombination rate in WD01 compared to our model for high $\gamma$ values.
On the contrary, with the same stellar effective temperature and the same gas temperature, BT94 find at $\gamma = 4\times 10^4$ $\mathbf{F}=\{0, 0.4, 0.6, 0.0\}$, and $\mathbf{F}=\{0.2, 0.5, 0.3, 0.0\}$, for planar and
spherical particles, respectively, resulting in lower ionized fractions compared to our model with $\mathbf{F}=\{0.4, 0.5, 0.1, 0\}$ (Fig.~\ref{fig_pop_fraction}).

\subsection{Heating efficiency}

\begin{figure}[!h]
    \centering
    \includegraphics[width=9.0cm]{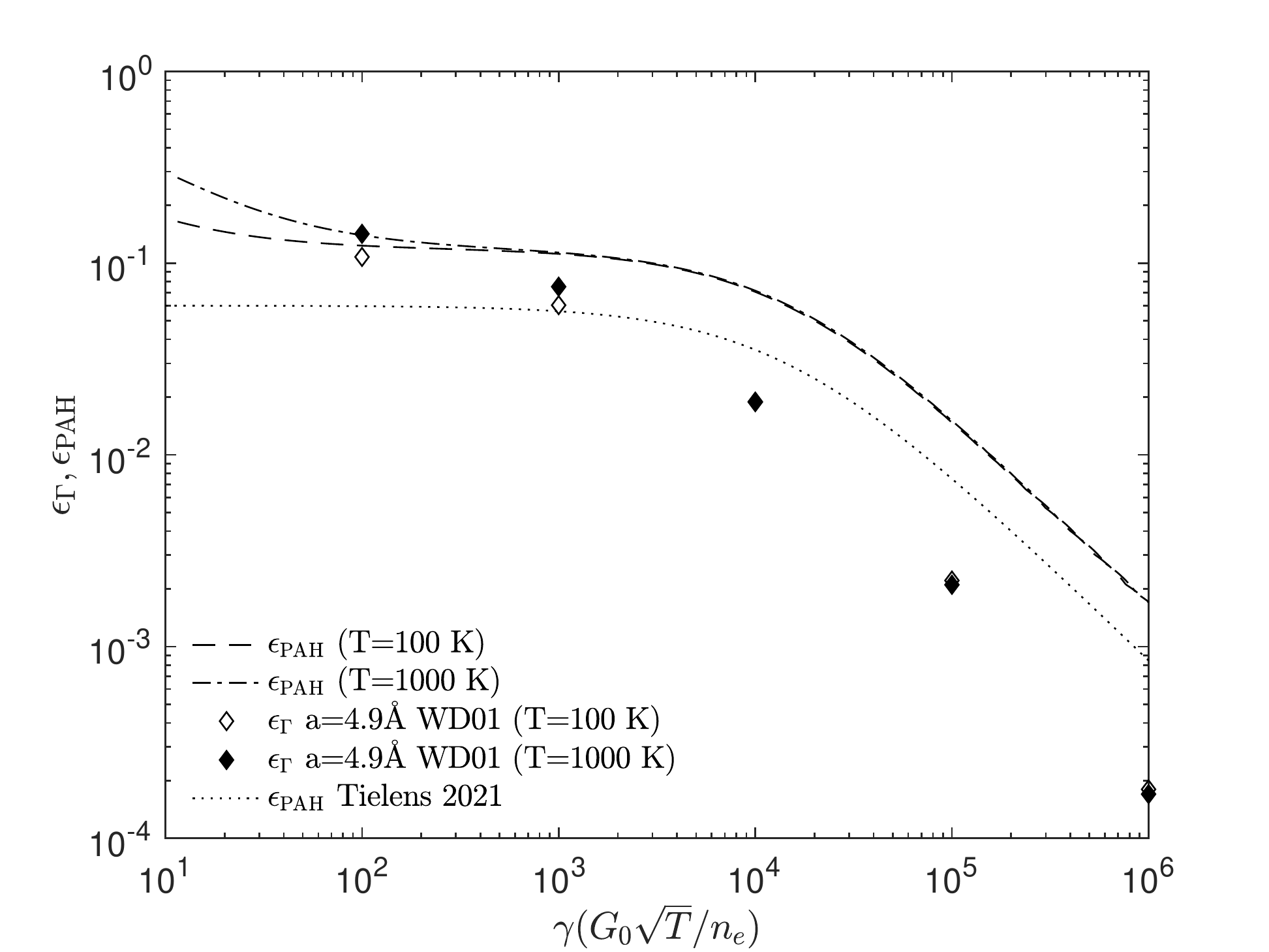}
    \caption{Photoelectric efficiency $\epsilon_{ PAH}$ computed for $N_{C}=54$, 
    $T_{\rm eff} = 30~000$\,K,
    and $T$=100\,K and 1000\,K with the PAH heating model. 
    For comparison, we give the photoelectric efficiency $\epsilon_{\Gamma}$ computed by WD01 for grains with a size $a=4.9$\AA~
    equivalent to $N_{C}=54$ (Eq.~\ref{eq_size_to_NC}), and the heating efficiency for PAHs given in \citet{tie21}.  }
    \label{fig_benchmark_efficiency}
\end{figure}

In Fig.~\ref{fig_benchmark_efficiency}, we present the results of 
the computation of $\epsilon_{\rm PAH}$ with the {PE heating model by PAHs}
for $T_{\rm eff} = 30~000$~K, $N_{\rm C}=54$, and two values of the gas temperature, $T=100$\,K and  $T=1000$\,K. 
We compare $\epsilon_{\rm PAH}$ with the efficiencies $\epsilon_{\Gamma}$ derived for the same physical conditions
by WD01, for grains of size $a=4.9$\,\AA~, corresponding to $N_{\rm C}=54$ (Eq.~\ref{eq_size_to_NC}). 
For $\gamma=100$ we find a good agreement between the two models, 
for both gas temperatures. 
For higher values of $\gamma$, $\epsilon_{\rm PAH}$ is systematically higher than 
$\epsilon_{\Gamma}$. The difference increases with increasing $\gamma$, up to about 
an order of magnitude for values of $\gamma \gtrsim 10^5$. 
This {can be rationalized} by a slower decrease of the ionization to 
recombination rate ratio with $\gamma$ in our model as compared to WD01 
(see previous section). 
\citet{tie21} provides the following formula for the PAH heating 
efficiency:
\begin{equation}
    \epsilon_{\rm PAH}= \frac{0.06}{1+7\times10^{-5}~\gamma},
    \label{eq_epsilon_pah_tielens}
\end{equation}
which we also display in Fig.~\ref{fig_benchmark_efficiency}. 
This curve is also below the heating efficiency we infer for PAHs.
At $\gamma \lesssim 10^3 $ this is due to the the absence of 
photo-detachment in Eq.~\ref{eq_epsilon_pah_tielens}.
For larger values of $\gamma \gtrsim 10^3$, the difference stems, notably, from the use of a photoelectron yield $Y(E)=0.3$  
at 10 eV by \citet{tie21}, while we use a value closer to $Y(E)=0.5$ (see Fig.~\ref{fig_yield}). 
Since BT94 do not provide the photoelectric heating efficiency 
curves for different sizes, we cannot directly compare our results to their model.

\subsection{Heating rates}

\begin{figure}[!h]
    \centering
    \includegraphics[width=9.0cm]{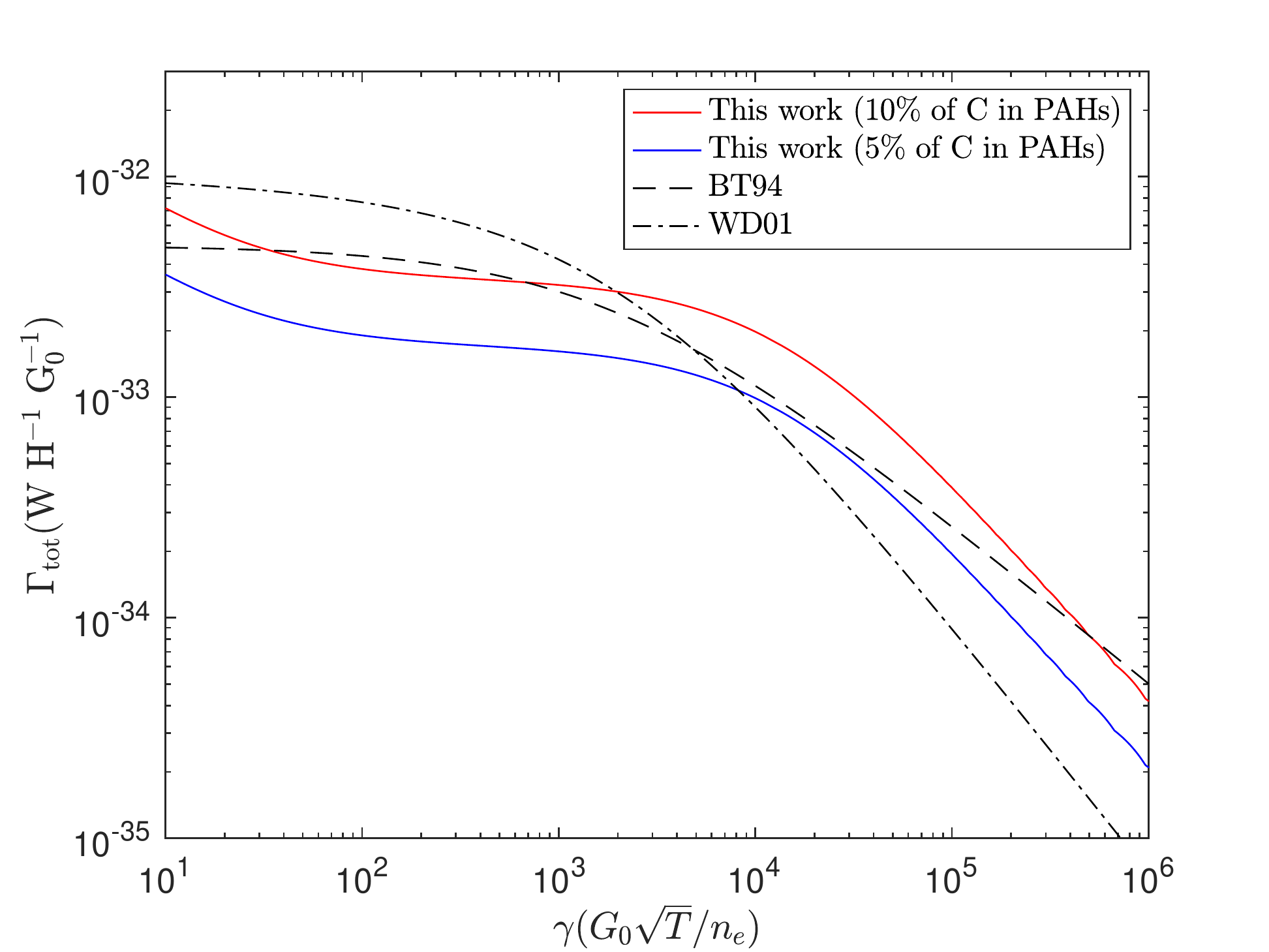}
    \caption{Photoelectric heating rate of the gas.
    Dash-dotted line:  average model from \citet{weingartner_Draine_01_sizedistrib}
    (see text for details). 
    Dashed line: model from \citet{bakesandtielens94}. Continuous lines:
    PAH model (this work) for two PAH abundances, that is $f_C=0.05, 0.1$.}
    \label{fig_benchmark_heatrate}
\end{figure}

\noindent The PAH heating rate we derived as a function of $\gamma$ is shown in Fig.~\ref{fig_benchmark_heatrate} for $T=100$\,K, $N_{C}=54$, $T_{\rm eff}=3\times10^4$\,K, and two PAH abundances. 
We compare these values to the total heating rate of WD01 averaged over several dust size distributions 
(Fig.~15 of WD01) and that of BT94, obtained for the same physical conditions.  
{Overall, the curves agree well. This implies that, with a purely molecular model, PAHs 
can produce a heating rate comparable to that of earlier models including a full size distribution 
of dust grains.} 
We can also compare the results of our model with those of \citet{Verstraete1990} for the  diffuse ISM. The physical conditions are
$G_0=1.7$ with the \citet{dra78} field, a gas temperature of $T=80$\,K, and an electron density $n_{\rm e}=3\times10^{-2}$\,cm$^{-3}$ yielding 
$\gamma \sim 500$\,K$^{1/2}$\,cm$^{-3}$.
With a value of $f_{\rm C}=0.1$, as in \citet{Verstraete1990},
our model yields $\Gamma_{\rm PAH}=2.7\times10^{-33}$\,W\,H$^{-1}$ (Tab. \ref{tab_diffuse-ism}). 
The latter authors find a value that is larger by a factor $\sim 3$, that is 
$\Gamma_{\rm PAH}= 8\times10^{-33}$\,W\,H$^{-1}$.
This is probably due to a combination of the following factors:  1) the higher ionization yield used by these authors, for example $Y \sim 0.7$ at 10 eV compared to $Y \sim 0.5$ in our model (Fig.~\ref{fig_yield}), 
2) their larger value for the partition coefficient $\langle \gamma_e (E) \rangle=0.5$ compared to 0.46 in our model, 
3) the fact that we include the ionization of cations in our model, 
which affects the charge balance and reduces the heating rate,
because the dications do not heat the gas.

\begin{table}[h!]
\centering
\caption{ Gas heating efficiency ($\epsilon_{\rm PAH}$) and heating rate ($\Gamma_{\rm PAH}$) for the diffuse ISM, computed with the PAH model.}
\begin{tabular}{ccc} 
\hline
$\epsilon_{\rm PAH}$ & $\Gamma_{\rm PAH}$ ($10^{-33}$W~H$^{-1}$) &  Radiation field  \\
\hline
\multicolumn{3}{c}{This work}\\
\hline
0.037 &  2.7 & \citet{dra78}  \\
0.059 & 1.6 &  \citet{Habing68}  \\
0.039 & 2.0 &  \citet{Mathis1983}  \\
\hline         
\end{tabular}
\tablefoot{Computed with the PAH model, for $T=80$ K, $f_{\rm C} =0.1$ and $n_e=3\times10^{-2}$, using three different model spectra of UV 
interstellar radiation field. 
\label{tab_diffuse-ism}}
\end{table}

\subsection{Empirical formulas for the heating efficiency and rate}

\begin{figure}[!ht]
    \centering
     \includegraphics[width=10cm]{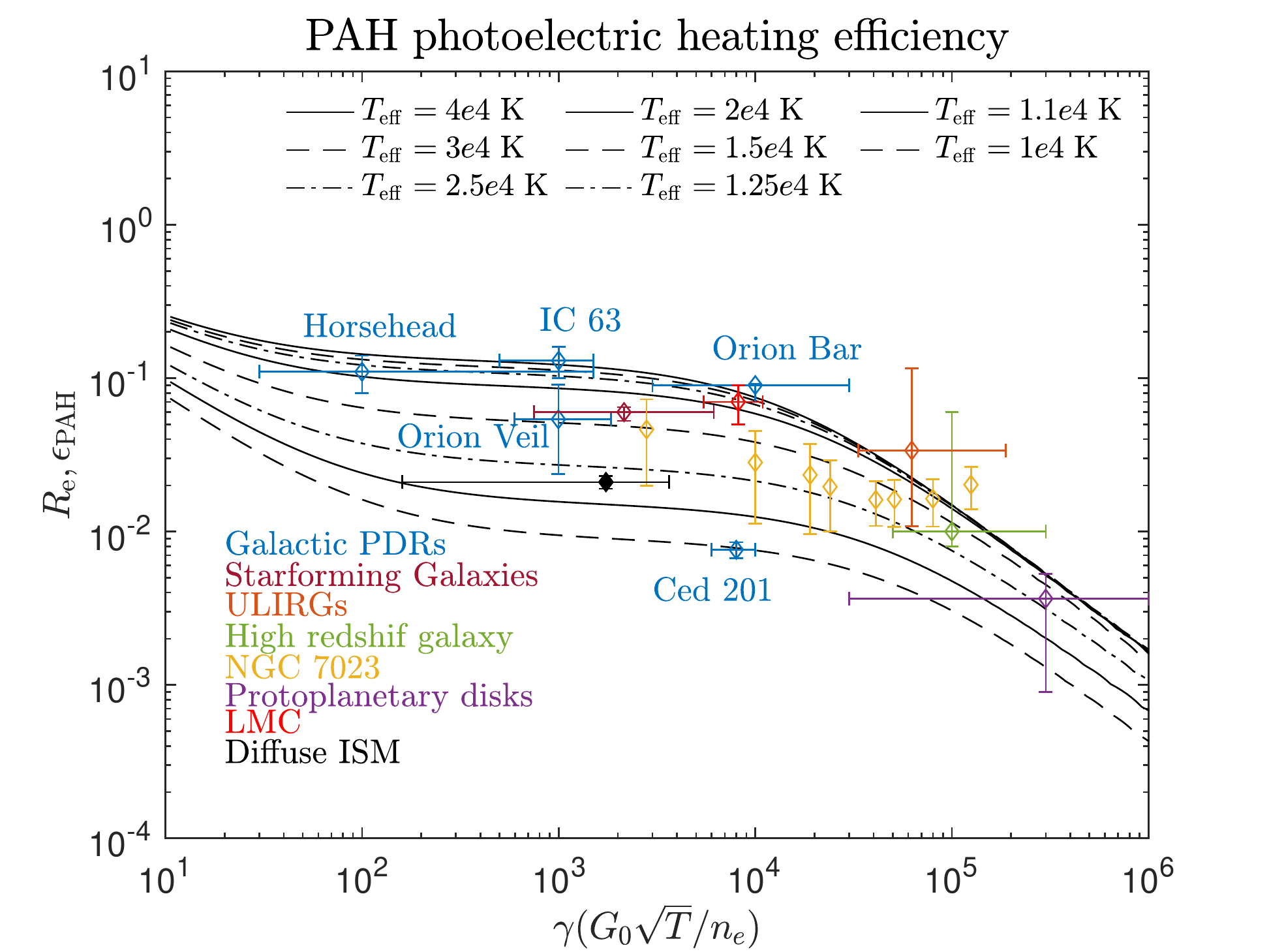}
    \caption{Observed emission ratio $R_{\rm e}$
    (Table~\ref{tab_all-objects}) and {modeled} photoelectric heating efficiency {for PAHs}, $\epsilon_{\rm PAH}$,
    as a function of the ionization parameter $\gamma$, for various stellar effective temperatures. 
    The upper curve corresponds to $T_{\rm eff} = 4\times10^4$K, and the lowest curve to $T_{\rm eff} = 10^4$K.
    All models use $N_C=54$ and a gas temperature of $T=500$\,K.}
\label{fig_all-objects-epsilon-vs-obs}
\end{figure}

We computed $\epsilon_{\rm PAH}$ using the molecular model
presented in Sect.~\ref{sec_model}, for a fixed size $N_{\rm C}=54$ (corresponding to cicumcoronene, see Table~\ref{tab_listemolecules}), 
and a gas temperature $T=500$\,K, using radiation fields with stellar spectral types
at various effective temperatures ($T_{\rm eff}$ in the $10^4-
4\times10^4$ K range). For $T_{\rm eff} \geq 3\times10^4$,
we used stellar spectra from the Pollux database~\footnote{\url{http://npollux.lupm.univ-montp2.fr}} 
\citep{pal10} computed with the CMFGEN code \citep{hil98}. The lower temperature 
stellar spectra are from \citet{Kurucz1993} database.
The resulting PE efficiency curves for PAHs as a function of the ionization 
parameter $\gamma = G_0\sqrt{T}/n_{\rm e}$ are shown in Fig.~\ref{fig_all-objects-epsilon-vs-obs}.
The values of $\epsilon_{\rm PAH}$ can be fitted using the following empirical formula:
\begin{equation}
\epsilon_{\rm PAH}
\approx \frac{a}{1+b \times \gamma}+
\frac{c}{1+d \times \gamma}.
\label{Eq_analytic_PE}
\end{equation}
This is similar to the expression, proposed by 
\citet{Tielens_book05} (see Eq.~\ref{eq_epsilon_pah_tielens})
but includes an additional term to take into account photo-detachment 
at low $\gamma$ values. 
We fit the analytical expression of Eq.~\ref{Eq_analytic_PE} to the 
model derived values $\epsilon_{\rm PAH}$ (Fig.~\ref{fig_all-objects-epsilon-vs-obs}) 
for all radiation fields, and provide the obtained values of 
the parameters in Table~\ref{tab_fit-paramters}.
Using Eq.~\ref{eq_Ptot} we computed $P_{\rm Rad}$ for $N_C=54$ 
for $T=500$K and $T_{\rm eff}$ between $10^4$ and $4\times10^4$K, as a function of $G_0$.
The result of this calculation, combined to Eq.~\ref{eq_heat_rate_model} using $ N_{\rm C} =54$, 
and $[C]=2.7\times10^{-4}$ per H atom \citep{tie21} yields an average heating rate for PAHs: 
\begin{equation}
    \Gamma_{\rm PAH} = 3.0 \pm 0.6 \times 10^{-31} \epsilon_{\rm PAH} G_0 f_{\rm C} ~~~\text{(W\,H$^{-1}$)},
    \label{eq_heating_rate}
\end{equation}
where the uncertainty on the numerical coefficient is the standard deviation among all numerical values obtained with the model.

\begin{table}[]
\centering
\caption{Parameters of the fit with analytical expression in Eq.~\ref{Eq_analytic_PE} to the model derived {curves of the photoelectric heating 
efficiency for PAHs}, $ \epsilon_{\rm PAH}$. \label{tab_fit-paramters}}
\begin{tabular}{cccccc} 
\hline
 $T_{\rm eff} (/10^{4})$ &  S. Type &  $a$  & $b$ & $c$ & $d (/10^{-5})$  \\
\hline
$4.00$ & O8 & 0.949 & 0.641 & 0.130 & 7.45 \\
$3.00$ & O9.5 & 0.947 &0.646 & 0.119 & 6.77  \\
$2.50$ & B0.5 & 0.948 & 0.645 & 0.108 & 6.20   \\
$1.50$ & B5 & 0.947 & 0.660 & 0.089 & 5.20  \\
$2.00$ &  B2.5 & 0.947 & 0.732 & 0.0516 & 3.54 \\
$1.50$ & B5 & 0.968 & 0.880 & 0.0267 & 2.60 \\
$1.25$ & B9 & 0.985 & 1.08 & 0.015 & 2.18 \\
$1.00$ & A0 & 0.992 & 1.35 & 0.00889 & 1.91 \\

\hline
\end{tabular}
\tablefoot{Fit results are provided for radiation fields of varying 
effective temperatures $T_{\rm eff}$ and corresponding spectral types.
All fits are computed on 1001 model points and yield 
a determination coefficient $R^2\approx 0.99$.}
\end{table}

 \section{Charge state and photo-electric heating of PAHs: Model vs observations}
\label{sec_model-vs-obs}

In this section we discuss the comparison between the observational 
results presented in Sect.~\ref{sec_observations} and the results of the molecular model described in Sect.~\ref{sec_model}.
More specifically, for the case of NGC 7023, we compare $R_{\rm i}$, 
the ionization fraction of PAHs derived from the {analysis of the mid-infrared spectra}  of the PDR (see Appendix~\ref{app_derivation_fion_obs_ngc7023}) to $f_i$, the modeled PAH ionization fraction. We also compare the emission ratio, $R_{\rm e}$, to {computed}  values of $\epsilon_{\rm PAH}$. 
If PAHs are the main source of gas heating, then the later two values and their variations should be in close agreement. {We then extend this comparison to}  our sample of galactic and extragalactic sources. 
{In addition}, for NGC 7023, the Orion Bar, and the diffuse ISM, we compare the total gas cooling $\Lambda_{\rm gas}$ to the modeled heating rate by PAHs $\Gamma_{\rm PAH}$.

\subsection{Comparison of model results with observations in NGC 7023}
\label{sec_ngc7023_obs_vs_mod}

\begin{figure}[!hb]
    \centering
     \includegraphics[width=\hsize]{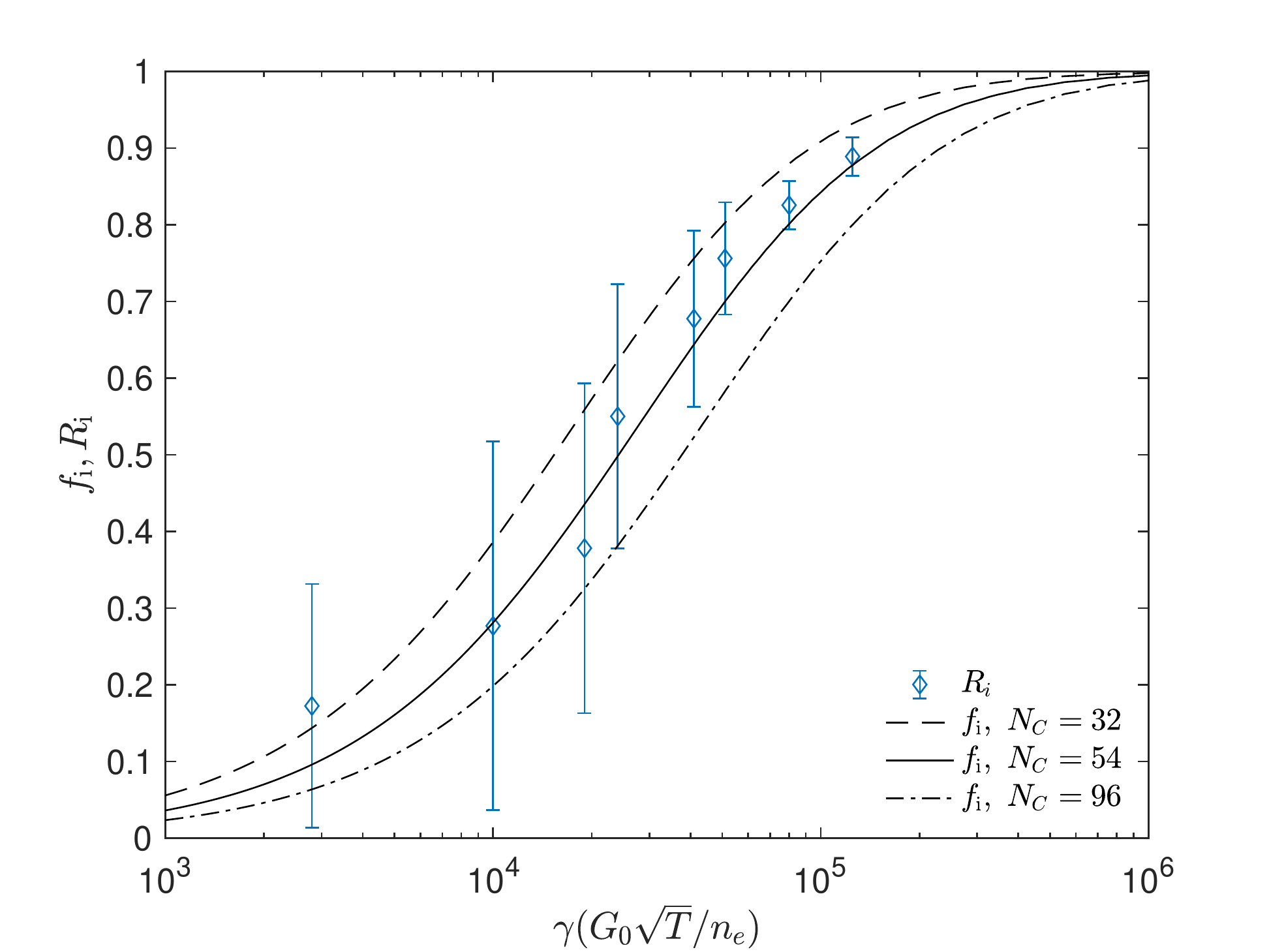}
    \caption{ Variation {with $\gamma$} of the PAH ionization fraction $R_{\rm i}$ 
    in NGC 7023 (diamonds with errobars).  {Values} derived from the map in
    Fig.~\ref{fig_maps-efficiency-ionization} for the eight angular distances 
between 20 and 55'' from HD 200775 given in Table~\ref{tab_parametresphysiques}.
    The values of $f_{\rm i}$ from the PAH model are shown with lines for a gas 
    temperature $T=500$K, and three PAH sizes.}
    \label{fig_ionization_fraction_7023}
\end{figure}

\begin{figure}[!ht]
    \centering
     \includegraphics[width=\hsize]{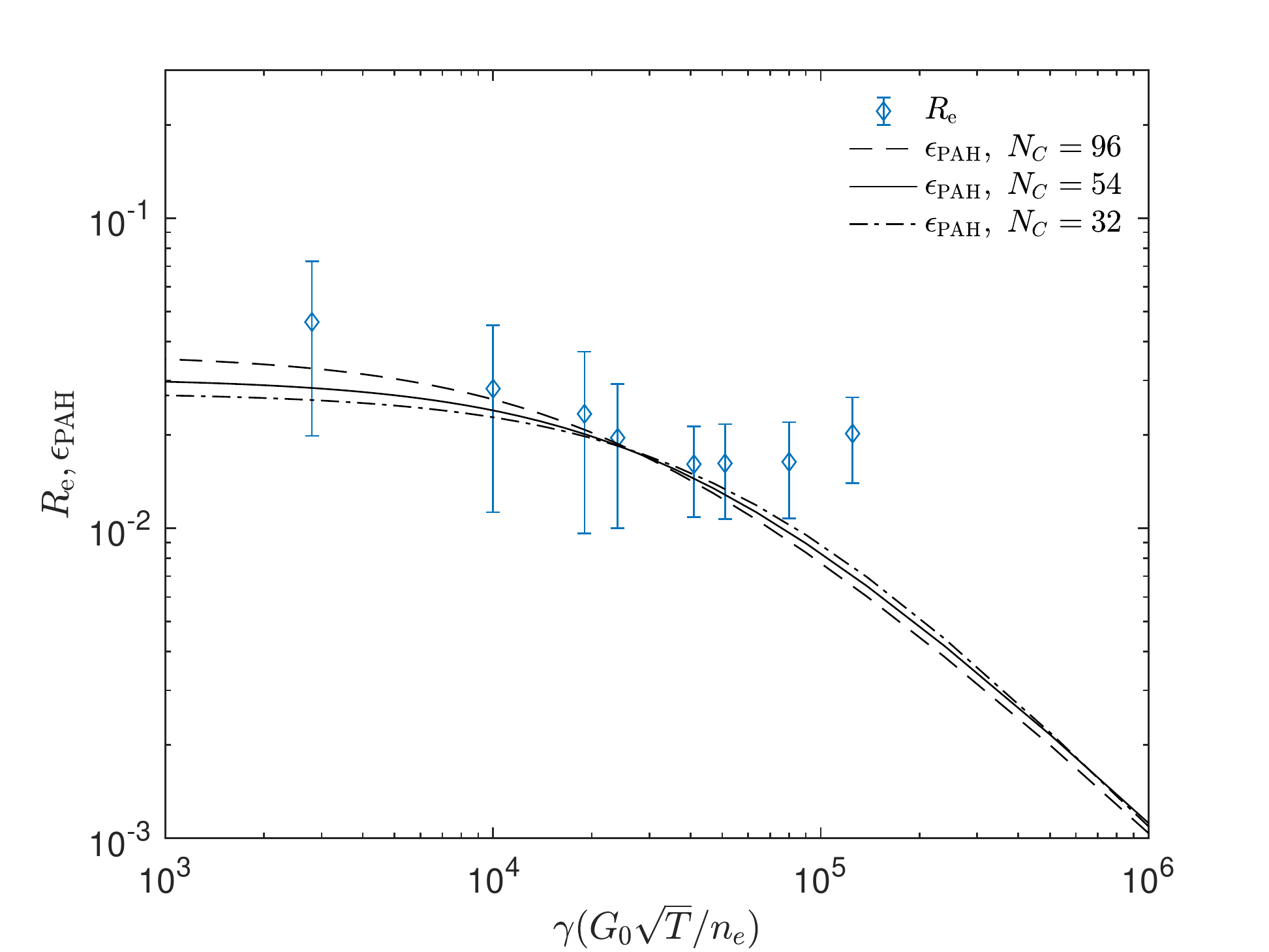}
    \caption{{ {Variation with $\gamma$ of the emission ratio} $R_{\rm e}$ in NGC 7023  derived from the map in
    Fig.~\ref{fig_maps-efficiency-ionization} for the eight angular distances 
between 20 and 55'' from HD 200775 given in Table~\ref{tab_parametresphysiques}.
The values of $\epsilon_{\rm PAH}$ from the PAH model are shown with lines for a gas 
temperature $T=500$K, and three PAH sizes.}}
    \label{fig_efficiency_7023}
\end{figure}

We first computed the model for three 
PAH sizes that is $N_C = 96,~54,~32$, using the {physical parameters in Tab.~\ref{tab_parametresphysiques}.} and the stellar spectrum of 
HD 200775 (see Appendix~\ref{app_derivation_gamma_NGC7023}).
The gas temperature in NGC\,7023~NW varies between 750 and 150\,K. We therefore adopt a fixed value of $T=500$\,K, since this parameter has only a small effect on 
the model results (Fig.~\ref{fig_benchmark_efficiency}).
The comparison of model results with the  observational diagnostics is provided for the {eight positions listed} in Table~\ref{tab_parametresphysiques}. 
In Fig.~\ref{fig_ionization_fraction_7023} we compare the ionization fraction obtained with the model, $f_{\rm i}$, to those derived 
from observations in NGC 7023, $R_{\rm i}$, as a function of $\gamma$. 
The {obtained good} agreement suggests that the model is able to 
{quantify}  the evolution of the charge state of PAHs in 
NGC\,7023. The PAH size plays a small role in {the values of $f_{\rm i}$,} but 
better results are obtained  for PAHs with $N_C \sim 54$. 
In Fig.~\ref{fig_efficiency_7023}, we compare  $R_{\rm e}$  as measured in NGC 7023 with values of the heating efficiency, $\epsilon_{PAH}$, derived from the model for three PAH sizes. The agreement in terms of curve shape and absolute values is good, with all observed values intersecting the model, except for the closest positions to the star. There are a number of parameters that 
could explain this discrepancy including less precise values for the physical conditions in this region, or a possible additional source of heating in this environment, such as shocks due to the expansion of a small HII region. This HII region could also contribute to some fraction of the [CII] emission that we include in the total cooling. Overall, however, $R_{\rm e}$ compares well to 
$\epsilon_{\rm PAH}$. 
Finally, we can also compare the total photoelectric heating rate {by PAHs}
$\Gamma_{\rm PAH}$ to the total gas cooling rate 
$\Lambda_{\rm gas} = 4\pi \times I_{\rm gas} / N_{\rm H}$, with 
$N_{\rm H}$ the total hydrogen column density. 
\citet{joblin_gas_2010} provide values for $N_{\rm H}$ in the NGC 7023
PDR, which are $N_{\rm H}=10.5, 8.1$ and $3.9\times 10^{21}$\,cm$^{-2}$, 
at 50, 48 and 47'' from the star, respectively.
We thus adopt $N_{\rm H}= 7.5 \pm 3.3 \times 10^{21}$\,cm$^{-2}$. 
At distances 45 - 50'' from the star, we derive 
$I_{\rm gas} = 3.8 \pm 1.6 \times 10^{-6}$\,W~m$^{-2}$~sr$^{-1}$. 
This yields a cooling rate $\Lambda_{\rm gas} = 6.4 \pm 4.6 \times 10^{-31}$\,W~H$^{-1}$
for this range of distances. 
{The heating rates by PAHs derived from the model} for the physical conditions at 45 and 50'' (Table~\ref{tab_parametresphysiques}) 
and a value $f_C=0.05$ are, for both positions, $\Gamma_{\rm PAH} = 6.5\times 10^{-31}$\,W~H$^{-1}$, 
in agreement with the cooling rates {derived from the observations}.

\subsection{Comparison of model results with observations for the sample of Galactic and extragalactic regions}

\subsubsection{Photoelectric heating efficiency}

Figure~\ref{fig_all-objects-epsilon-vs-obs}  compares the variation with $\gamma$ of the observational diagnostic $R_{\rm e}$  
to the modeled values of $\epsilon_{\rm PAH}$ {for the set of objects listed in Table~\ref{tab_all-objects}}. 
Here we adopt $N_C=54$ in the model, and a gas temperature $T=500$\,K, but we note that the choice of values for these 
parameters has a limited effect on the model results for $\gamma \gtrsim 10^3$  (see Figs.~\ref{fig_ionization_fraction_7023}, \ref{fig_efficiency_7023},
and \ref{fig_benchmark_efficiency}), applicable to most sources.  {In contrast,} since the effective temperature of the 
radiation field has strong effects on the value of $\epsilon_{\rm PAH}$ (see \citealt{spa94}), 
we compute the models with appropriate effective temperatures. 
Figure~\ref{fig_all-objects-epsilon-vs-obs} shows that, in general, there 
is a good agreement between $R_{\rm e}$ and
$\epsilon_{\rm PAH}$, in terms of {curve shape}  but also of absolute values.  
The upper part of Fig.~\ref{fig_all-objects-epsilon-vs-obs} presents the comparison 
between $R_{\rm e}$ and $\epsilon_{\rm PAH}$ for interstellar regions 
irradiated by the most massive stars with $20 000 \leq T_{\rm eff} \leq 40 000$\,K. 
In the Orion Veil, $R_{\rm e}$ is somewhat lower than the predicted 
$\epsilon_{\rm PAH}$, for a star with $T_{\rm eff} = 40~000$~K (which
is appropriate for $\Theta^1$ Orionis C). This lower value can be explained by 
the fact that the radiation field {impinging on} the Veil is likely softer than that of 
$\Theta^1$ Orionis C, because of the UV absorption by dust and PAHs 
situated in between the Veil and the massive star. The Orion Bar which is much closer 
to $\Theta^1$ Orionis C has a value of $R_{\rm e}$ which is in
good agreement with that of $\epsilon_{\rm PAH}$ for $T_{\rm eff} = 40~000$~K. 
The other nebulae 
illuminated by O or early B stars in the sample, that is the Horsehead and IC 63 also 
have values of $R_{\rm e}$ in good agreement with   {modeled values of} $\epsilon_{\rm PAH}$ for $T_{\rm eff} = 30~000 - 40~000$~K.
For the LMC, $R_{\rm e}$ is in good agreement with the 
$\epsilon_{\rm PAH}$ curve for the most massive stars 
($T_{\rm eff} > 3\times10^4$K). This is compatible with 
the fact that the values we use to derive  $R_{\rm e}$ from
\citet{rub09} and \citet{leb12} concern star-forming regions. 
 $R_{\rm e}$ in star-forming galaxies falls within the range of the models with 
$T_{\rm eff} \sim 25~000$~K, implying an average radiation field which is 
softer than that of O stars and dominated by B-type stars. 
$R_{\rm e}$ for ULIRGs is also in good agreement with 
the model. ULIRGs have a lower heating efficiency as compared to star-forming galaxies 
of the local universe, which is well explained in the model 
by the higher value of the ionization parameter $\gamma$ in these galaxies 
due to intense {massive} star formation, increasing the average integrated intensity
of the radiation field $G_{0}$ \citep{dia17,mck21}. 
The $R_{\rm e}$ value of the high redshift galaxy 
GS IRS20 is also in good agreement with the model. 
 Figure~\ref{fig_all-objects-epsilon-vs-obs} also presents the comparison 
between $R_{\rm e}$ and $\epsilon_{\rm PAH}$ for interstellar regions 
irradiated by intermediate-mass stars with $10 000 \leq T_{\rm eff} \leq 15 000$\,K.
This include NGC 7023 which was presented in detail in Sect.~\ref{sec_ngc7023_obs_vs_mod}.
The case of Ced 201 is particularly interesting, because it has a low PAH heating
efficiency, which is well explained by the fact that this nebula is illuminated 
by a B 9.5 star, whose cooler spectrum is less efficient to ionize PAHs (see Sect.~\ref{app_gamma}). 
The value of $R_{\rm e}$ for Ced 201 indeed falls close to the model 
value of $\epsilon_{\rm PAH}$ for a $T_{\rm eff} = 10~000$~K, while a B 9.5 spectral type
has $T_{\rm eff} \sim 11 000$~K. 
The lowest efficiencies are observed in protoplanetary disks
(Fig.~\ref{fig_all-objects-epsilon-vs-obs}).
In this case too there is a good agreement between $R_{\rm e}$ and the modeled
value of $\epsilon_{\rm PAH}$, which are low due to both the high $\gamma$ values and cool spectral types (A and B) for these sources. 
Table~\ref{tab_diffuse-ism} gives the values derived for $\epsilon_{\rm PAH}$ for diffuse ISM conditions, 
that is $T=80\,K$ and $n_e=3\times10^{-2}$\,cm$^{-3}$ for three classical radiation fields. Observations yield $R_{\rm e}=2.1\%$
(Table~\ref{tab_all-objects} and Appendix~\ref{app_derivation_gamma_all_objects}),
close but about a factor of $\sim$ 2-3 lower than the model
values, that is $\epsilon_{\rm PAH} = 3.7, 5.9$ and $ 3.9 \%$ for the interstellar radiation fields of \citet{draine1978}, \citet{Mathis1983} and \citet{Habing68}, respectively.

\subsubsection{Heating rates}

We compute the heating rates in sources for which there is a good estimate of the column density of the cooling gas, that is the Orion Bar \citep{salgado_orion_map_eps_2016} and a diffuse ISM line of sight \citep{bou96}.
For the Orion bar, with $I_{\rm gas}=6.6\times 10^{-5}$ W~m$^{-2}$ (Appendix\ref{app_derivation_gamma_all_objects}) and $N_{\rm H} = 7 \times 10^{22}$ cm$^{-2}$ \citep{salgado_orion_map_eps_2016},
$\Lambda_{\rm gas}=1.2 \times 10^{-30}$ W~H$^{-1}$. 
This is close to the modeled gas heating rate by PAHs, that is 
$\Gamma_{\rm PAH} = 4.2 \pm 2.0 \times 10^{-30}$ W~H$^{-1}$, 
considering $G_0=2\times10^4$ \citep{Joblin2018} and $f_C=0.01$ \citep{castellanos_c_2014_fc} and the range of $\gamma$
values in Table~\ref{tab_all-objects}.
For the diffuse ISM, \citet{Boulanger1998} find a cooling rate of $\Lambda_{\rm gas}= 3 \times 10^{-33}$~W~H$^{-1}$ from FIRAS 
observations, in good agreement with the model values of $\Gamma_{\rm PAH}$
presented in Table~\ref{tab_diffuse-ism}.

\section{Discussion}
\label{sec_discussion}

\subsection{The role of PAHs in neutral gas heating}

\label{sec_dominant-role-pahs}

The comparison presented in Sect.~\ref{sec_model-vs-obs} shows that
there is a good agreement between {the values of} $\epsilon_{\rm PAH}$ derived from 
the PAH heating model and {the values of the observational diagnostics} $R_{\rm e}$, 
for a wide range of physical conditions and environments. {In addition, there is} an excellent agreement in terms of the {spatial} variations of the charge state of PAHs in NGC 7023.
The agreement between the 
{modeled heating rates by PAHs} and the gas cooling rates {derived from the observations} is also good for 
NGC~7023, the Orion Bar, and the diffuse ISM. 
Overall, this suggests that photoelectric heating by 
PAHs  alone can explain the neutral gas heating in a number of environments.
While this has been suggested by several studies (see Introduction),
a direct assessment comparing the results from a molecular model
including state-of-the-art molecular parameters for PAHs to 
observational diagnostics, had not been conducted before. 
BT94 pointed out the importance of PAHs, however $~50\%$ of the gas heating 
in their model is due to grains larger than $N_C=1500$, while the present PAH 
model contains only PAHs of $N_C=54$~C atoms. 
This stresses the central role that PAHs play in the physics of the gas heating
of galaxies. 

\subsection{Heating by dust vs heating by PAHs}

\citet{pab21} and earlier works have compared the theoretical photoelectric heating efficiency of 
dust from BT94, $\epsilon$, to $R_{\rm IR} = \frac{I_{\rm gas}}{I_{\rm IR}}$, where ${I_{\rm IR}}$
is the total power radiated by dust in the far infrared. They found that $\epsilon > R_{\rm IR}$ 
by a factor $\sim 3$ or more. This discrepancy is likely due to the fact 
that $R_{\rm IR}$ cannot easily be compared to $\epsilon$, 
for two reasons. The first one is that, in the definition of BT94, $\epsilon$ is computed 
by integrating the power absorbed (emitted) by dust in the 6-13.6 eV range, while some 
fraction of the observed radiated power ${I_{\rm IR}}$ is in reality due to absorption 
of UV photons with energies below 6 eV, and therefore ${I_{\rm IR}}$ {overestimates the denominator value.} 
Still, it appears unlikely that this effect only can explain 
this factor of 3 difference, because most of the energy absorbed by dust does lie above 6 eV. 
{A most likely reason is} that ${I_{\rm IR}}$ includes an important fraction of emission from dust grains 
which do not contribute at all to photoelectric heating.

\subsection{PAH heating vs X-rays, cosmic rays, shocks, and turbulence in neutral gas.}

Since the gas heating rate by PAHs $\Gamma_{\rm PAH}$ 
depends on their abundance (Eq.~\ref{eq_heating_rate}),
in regions where they are depleted, such as dense molecular clouds where PAHs {are in condensed form}  \citep{Rapacioli2006}, other mechanisms are likely to take over, such as 
heating by cosmic rays (e.g. \citealt{padovani2009cosmic}). 
In regions where the UV radiation field is weak compared to other sources of 
radiation such as X-rays (e.g. in the inner regions of protoplanetary disks 
near T-Tauri stars, or near active galactic nuclei), heating by PAHs may also become negligible.
In PDRs, heating by PAHs is likely to dominate, but heating by H$_2$ can also become important
in dense regions near the dissociation front \citep{Bron2014}.
In some regions, shocks may become an important source 
of neutral gas heating. 
The competition between heating by PAHs and other sources of gas heating 
can be tested using $R_{\rm e}$: when this ratio is larger 
than the $\epsilon_{\rm{PAH}}$ values predicted by the model, this is indicative 
that some other processes are at play. 
For instance, shock heating of neutral gas at the scale of galaxies results in extreme values of 
$R_{\rm e}$ (e.g. >0.3, \citealt{app13}, Table 1), well above the maximum 
efficiencies derived with our model ($\epsilon_{\rm{PAH}}^{\rm Max}\sim 0.11$, 
see Table~\ref{tab_fit-paramters}). 
This illustrates how the PAH model efficiencies  can be used as a diagnostic of UV radiative feedback vs other sources of heating in galaxies.

\section{Conclusion}
\label{sec_conclusion}

In this article, we re-evaluate the contribution of PAHs to the photoelectric heating of neutral gas in astrophysical environments, using {a simple analytical model in which} state-of-the-art molecular parameters from laboratory measurements and quantum chemical calculations {are included}.
We find that, for standard abundances of PAHs, the heating rates {resulting from photoelectric heating by} PAHs alone is comparable to those produced using a full size distribution of dust grains in the classical models of \citet{bakesandtielens94} and \citet{weingartner_photoelectric_2001}. The modeled and observed charge states of PAHs and their variation with physical conditions in NGC 7023 are in excellent agreement. The {values of the} photoeletric heating efficiency derived for PAHs from the model for a wide range of Galactic and extragalatic regions are in good 
agreement with {the values of} the observational diagnostics $R_{\rm e}$, that is the ratio of gas emission to the sum of gas + PAH emission. The {values of}  the heating rate {by PAHs} derived from the model are also in good agreement with the observed gas cooling rates for the NGC 7023 nebula, the Orion Bar, and the diffuse ISM. 
Overall, this allows us to conclude that PAHs can explain most of the heating of the neutral gas in a variety of astrophysical environments, ranging from protoplanetary disks around young intermediate-mass stars to starburst galaxies in the early Universe. {Our study} highlights the importance to implement a robust description of the PE heating by PAHs in astrophysical codes computing the thermal balance of neutral gas, such as PDR models.  The {formalism} presented here is included in a simple analytical model  but it relies on {our knowledge of the molecular properties of PAHs. This knowledge is crucial and laboratory studies (experiments and quantum chemical calculations must be guided by astronomical observations.} 
In particular, the coming spectroscopic data obtained from the MIRI and NIRspec spectrometers {on board the James Webb Space Telescope} will provide unprecedented details {on the PAH populations \citep*{ber22}}.

\begin{acknowledgements}

OB acknowledges C. Pabst and J. R. Goicoechea for providing the data 
on the PE heating efficiency in Orion, J. Mc Kinney
for providing the data regarding the PE heating efficiency in galaxies, F. Boulanger for fruitful discussions on the heating and cooling of the diffuse ISM, E. Josselin and A. Palacios for their help on extracting the stellar 
spectra used in this paper, and I. Schroetter for support on the writing of the code.
The authors wish to thank L. Verstraete and V. Guillet for their 
comments on an early version of this {work.}
We are grateful to the anonymous referee for his detailed 
report which greatly improved the manuscript. 
{Finally, the research leading to these results has received funding from the European Research Council under the European Union’s Seventh Framework Programme (FP/2007-2013) ERC-2013-SyG, Grant agreement N$^{o}$610256 NANOCOSMOS. It has also been supported} by the Programme National “Physique et Chimie du Milieu 
Interstellaire” (PCMI) of CNRS/INSU with INC/INP co-funded by CEA and CNES, 
and through an APR grant provided by CNES.

\end{acknowledgements}

\bibliographystyle{aa} 
\bibliography{aanda} 

\begin{thebibliography}{91}
\expandafter\ifx\csname natexlab\endcsname\relax\def\natexlab#1{#1}\fi

\bibitem[{Acke \& van~den Ancker(2004)}]{ack04}
Acke, B. \& van~den Ancker, M.~E. 2004, \aap, 426, 151

\bibitem[{Alecian {et~al.}(2008)Alecian, Catala, Wade, Donati, Petit,
  Landstreet, B{\"o}hm, Bouret, Bagnulo, Folsom,
  {et~al.}}]{alecian2008characterization}
Alecian, E., Catala, C., Wade, G., {et~al.} 2008, Monthly Notices of the Royal
  Astronomical Society, 385, 391

\bibitem[{Allamandola {et~al.}(1985)Allamandola, Tielens, \&
  Barker}]{allamandola_polycyclic_1985}
Allamandola, L.~J., Tielens, A. G. G.~M., \& Barker, J.~R. 1985, \apj, 290, L25

\bibitem[{{Andrews} {et~al.}(2016){Andrews}, {Candian}, \&
  {Tielens}}]{Andrews2016}
{Andrews}, H., {Candian}, A., \& {Tielens}, A.~G.~G.~M. 2016, \aap, 595, A23

\bibitem[{Andrews {et~al.}(2018)Andrews, Peeters, Tielens, \& Okada}]{and18}
Andrews, H., Peeters, E., Tielens, A., \& Okada, Y. 2018, \aap, 619, A170

\bibitem[{{Appleton} {et~al.}(2013){Appleton}, {Guillard}, {Boulanger},
  {Cluver}, {Ogle}, {Falgarone}, {Pineau des For{\^e}ts}, {O'Sullivan}, {Duc},
  {Gallagher}, {Gao}, {Jarrett}, {Konstantopoulos}, {Lisenfeld}, {Lord}, {Lu},
  {Peterson}, {Struck}, {Sturm}, {Tuffs}, {Valchanov}, {van der Werf}, \&
  {Xu}}]{app13}
{Appleton}, P.~N., {Guillard}, P., {Boulanger}, F., {et~al.} 2013, \apj, 777,
  66

\bibitem[{{Bakes} \& {Tielens}(1994)}]{bakesandtielens94}
{Bakes}, E.~L.~O. \& {Tielens}, A.~G.~G.~M. 1994, \apj, 427, 822

\bibitem[{Bauschlicher {et~al.}(2018)Bauschlicher, Ricca, Boersma, \&
  Allamandola}]{bauschlicher_nasa_2018}
Bauschlicher, C.~W., Ricca, A., Boersma, C., \& Allamandola, L.~J. 2018, \apjs,
  234, 32

\bibitem[{{Benisty} {et~al.}(2013){Benisty}, {Perraut}, {Mourard}, {Stee},
  {Lima}, {Le Bouquin}, {Borges Fernandes}, {Chesneau}, {Nardetto},
  {Tallon-Bosc}, {McAlister}, {Ten Brummelaar}, {Ridgway}, {Sturmann},
  {Sturmann}, {Turner}, {Farrington}, \& {Goldfinger}}]{Benisty2013}
{Benisty}, M., {Perraut}, K., {Mourard}, D., {et~al.} 2013, \aap, 555, A113

\bibitem[{Bernard-Salas {et~al.}(2012)Bernard-Salas, Habart, Arab, Abergel,
  Dartois, Martin, Bontemps, Joblin, White, Bernard, \&
  Naylor}]{bernard-salas_spatial_orion_2012}
Bernard-Salas, J., Habart, E., Arab, H., {et~al.} 2012, \aap, 538, A37

\bibitem[{Bernard-Salas {et~al.}(2015)Bernard-Salas, Habart, Köhler, Abergel,
  Arab, Lebouteiller, Pinto, van~der Wiel, White, \&
  Hoffmann}]{bernard-salas_spatial_NGC7023_2015}
Bernard-Salas, J., Habart, E., Köhler, M., {et~al.} 2015, \aap, 574, A97

\bibitem[{Bern{\'e} {et~al.}(2022)Bern{\'e}, Habart, Peeters, \& the
  PDRs4All~Team}]{ber22}
Bern{\'e}, O., Habart, {\'E}., Peeters, E., \& the PDRs4All~Team. 2022,
  Publications of the Astronomical Society of the Pacific, 134, 054301

\bibitem[{{Bern{\'e}} {et~al.}(2015){Bern{\'e}}, {Montillaud}, \&
  {Joblin}}]{Berne2015}
{Bern{\'e}}, O., {Montillaud}, J., \& {Joblin}, C. 2015, \aap, 577, A133

\bibitem[{{Bern{\'e}} {et~al.}(2013){Bern{\'e}}, {Mulas}, \&
  {Joblin}}]{Berne2013}
{Bern{\'e}}, O., {Mulas}, G., \& {Joblin}, C. 2013, \aap, 550, L4

\bibitem[{Berné {et~al.}(2007)Berné, Joblin, Deville, Smith, Rapacioli,
  Bernard, Thomas, Reach, \& Abergel}]{berne_analysis_2007}
Berné, O., Joblin, C., Deville, Y., {et~al.} 2007, \aap, 469, 575

\bibitem[{Boersma {et~al.}(2016)Boersma, Bregman, \&
  Allamandola}]{boersma_charge_2016}
Boersma, C., Bregman, J., \& Allamandola, L.~J. 2016, \apj, 832, 51

\bibitem[{Boersma {et~al.}(2013)Boersma, Bregman, \&
  Allamandola}]{boersma_properties_2013}
Boersma, C., Bregman, J.~D., \& Allamandola, L.~J. 2013, \apj, 769, 117

\bibitem[{Boulais(2013)}]{boulais_stage}
Boulais, A. 2013, Méthodes de séparation aveugle de source et application à
  l'imagerie hyperspectrale en astrophysique

\bibitem[{{Boulanger}(1999)}]{bou99}
{Boulanger}, F. 1999, in Solid Interstellar Matter: The ISO Revolution, 20

\bibitem[{Boulanger {et~al.}(1996)Boulanger, Abergel, Bernard, Burton, Desert,
  Hartmann, Lagache, \& Puget}]{bou96}
Boulanger, F., Abergel, A., Bernard, J.-P., {et~al.} 1996, \aap, 312, 256

\bibitem[{{Boulanger} {et~al.}(1998){Boulanger}, {Boisssel}, {Cesarsky}, \&
  {Ryter}}]{Boulanger1998}
{Boulanger}, F., {Boisssel}, P., {Cesarsky}, D., \& {Ryter}, C. 1998, \aap,
  339, 194

\bibitem[{Bron(2014)}]{Bron2014}
Bron, E. 2014, PhD thesis, Universit{\'e} Paris Diderot

\bibitem[{Bréchignac {et~al.}(2014)Bréchignac, Garcia, Falvo, Joblin, Kokkin,
  Bonnamy, Parneix, Pino, Pirali, Mulas, \&
  Nahon}]{brechignac_photoionization_2014}
Bréchignac, P., Garcia, G.~A., Falvo, C., {et~al.} 2014, The Journal of
  Chemical Physics, 141, 164325

\bibitem[{Carelli {et~al.}(2013)Carelli, Grassi, \& Gianturco}]{car13}
Carelli, F., Grassi, T., \& Gianturco, F. 2013, Astronomy \& Astrophysics, 549,
  A103

\bibitem[{Castellanos {et~al.}(2014)Castellanos, Berné, Sheffer, Wolfire, \&
  Tielens}]{castellanos_c_2014_fc}
Castellanos, P., Berné, O., Sheffer, Y., Wolfire, M.~G., \& Tielens, A.~G.
  2014, \apj, 794, 83

\bibitem[{{Chokshi} {et~al.}(1988){Chokshi}, {Tielens}, {Werner}, \&
  {Castelaz}}]{Chokshi88}
{Chokshi}, A., {Tielens}, A.~G.~G.~M., {Werner}, M.~W., \& {Castelaz}, M.~W.
  1988, \apj, 334, 803

\bibitem[{Compi{\`e}gne {et~al.}(2011)Compi{\`e}gne, Verstraete, Jones,
  Bernard, Boulanger, Flagey, Le~Bourlot, Paradis, \& Ysard}]{com11}
Compi{\`e}gne, M., Verstraete, L., Jones, A., {et~al.} 2011, \aap, 525, A103

\bibitem[{de~Jong(1977)}]{dej77}
de~Jong, T. 1977, \aap, 55, 137

\bibitem[{D{\'\i}az-Santos {et~al.}(2017)D{\'\i}az-Santos, Armus, Charmandaris,
  Lu, Stierwalt, Stacey, Malhotra, Van Der~Werf, Howell, Privon,
  {et~al.}}]{dia17}
D{\'\i}az-Santos, T., Armus, L., Charmandaris, V., {et~al.} 2017, \apj, 846, 32

\bibitem[{Draine(1978)}]{dra78}
Draine, B.~T. 1978, \apjs, 36, 595

\bibitem[{{Draine}(1978)}]{draine1978}
{Draine}, B.~T. 1978, \apjs, 36, 595

\bibitem[{d’Hendecourt \& L{\'e}ger(1987)}]{dhe87}
d’Hendecourt, L. \& L{\'e}ger, A. 1987, Astron. Astrophys, 180, L9

\bibitem[{{Finkenzeller}(1985)}]{Finkenzeller85_HD200775}
{Finkenzeller}, U. 1985, \aap, 151, 340

\bibitem[{{Fuente} {et~al.}(1996){Fuente}, {Martin-Pintado}, {Neri}, {Rogers},
  \& {Moriarty-Schieven}}]{Fuente1996_filament}
{Fuente}, A., {Martin-Pintado}, J., {Neri}, R., {Rogers}, C., \&
  {Moriarty-Schieven}, G. 1996, \aap, 310, 286

\bibitem[{Habart {et~al.}(2005)Habart, Abergel, Walmsley, Teyssier, \&
  Pety}]{hab05}
Habart, E., Abergel, A., Walmsley, C., Teyssier, D., \& Pety, J. 2005, \aap,
  437, 177

\bibitem[{{Habing}(1968)}]{Habing68}
{Habing}, H.~J. 1968, \bain, 19, 421

\bibitem[{Helou {et~al.}(2001)Helou, Malhotra, Hollenbach, Dale, \&
  Contursi}]{hel01}
Helou, G., Malhotra, S., Hollenbach, D.~J., Dale, D.~A., \& Contursi, A. 2001,
  \apj Letters, 548, L73

\bibitem[{Hillier \& Miller(1998)}]{hil98}
Hillier, D.~J. \& Miller, D.~L. 1998, \apj, 496, 407

\bibitem[{Iida {et~al.}(2021)Iida, Kuma, Kuriyama, Furukawa, Kusunoki, Tanuma,
  Hansen, Shiromaru, \& Azuma}]{Iida22}
Iida, S., Kuma, S., Kuriyama, M., {et~al.} 2021, Phys. Rev. A, 104, 043114

\bibitem[{{Joblin} {et~al.}(2018){Joblin}, {Bron}, {Pinto}, {Pilleri}, {Le
  Petit}, {Gerin}, {Le Bourlot}, {Fuente}, {Berne}, {Goicoechea}, {Habart},
  {K{\"o}hler}, {Teyssier}, {Nagy}, {Montillaud}, {Vastel}, {Cernicharo},
  {R{\"o}llig}, {Ossenkopf-Okada}, \& {Bergin}}]{Joblin2018}
{Joblin}, C., {Bron}, E., {Pinto}, C., {et~al.} 2018, \aap, 615, A129

\bibitem[{Joblin {et~al.}(2010)Joblin, Pilleri, Montillaud, Fuente, Gerin,
  Berné, Ossenkopf, Le~Bourlot, Teyssier, Goicoechea, Le~Petit, Röllig,
  Akyilmaz, Benz, Boulanger, Bruderer, Dedes, France, Güsten, Harris, Klein,
  Kramer, Lord, Martin, Martin-Pintado, Mookerjea, Okada, Phillips, Rizzo,
  Simon, Stutzki, van~der Tak, Yorke, Steinmetz, Jarchow, Hartogh, Honingh,
  Siebertz, Caux, \& Colin}]{joblin_gas_2010}
Joblin, C., Pilleri, P., Montillaud, J., {et~al.} 2010, \aap, 521, L25

\bibitem[{Joblin {et~al.}(2020)Joblin, Wenzel, Castillo, Simon, Sabbah,
  Bonnamy, Toublanc, Mulas, Ji, Giuliani, \& Nahon}]{Joblin2020}
Joblin, C., Wenzel, G., Castillo, S.~R., {et~al.} 2020, J. Phys. Conf. Ser.,
  1412, 062002

\bibitem[{{Jochims} {et~al.}(1996){Jochims}, {Baumgaertel}, \&
  {Leach}}]{Jochims1996}
{Jochims}, H.~W., {Baumgaertel}, H., \& {Leach}, S. 1996, \aap, 314, 1003

\bibitem[{Jura(1976)}]{jur76}
Jura, M. 1976, \apj, 204, 12

\bibitem[{{Kurucz}(1993)}]{Kurucz1993}
{Kurucz}, R.~L. 1993, VizieR Online Data Catalog, VI/39

\bibitem[{Le~Petit {et~al.}(2006)Le~Petit, Nehme, Le~Bourlot, \&
  Roueff}]{le_petit_model_2006}
Le~Petit, F., Nehme, C., Le~Bourlot, J., \& Roueff, E. 2006, \apjs, 164, 506

\bibitem[{{Lebouteiller} {et~al.}(2012){Lebouteiller}, {Cormier}, {Madden},
  {Galliano}, {Indebetouw}, {Abel}, {Sauvage}, {Hony}, {Contursi}, {Poglitsch},
  {R{\'e}my}, {Sturm}, \& {Wu}}]{leb12}
{Lebouteiller}, V., {Cormier}, D., {Madden}, S.~C., {et~al.} 2012, \aap, 548,
  A91

\bibitem[{{Leger} {et~al.}(1989){Leger}, {D'Hendecourt}, \&
  {Defourneau}}]{Leger1989_PAHmodel}
{Leger}, A., {D'Hendecourt}, L., \& {Defourneau}, D. 1989, \aap, 216, 148

\bibitem[{{Leger} \& {Puget}(1984)}]{leger_puget84}
{Leger}, A. \& {Puget}, J.~L. 1984, \aap, 500, 279

\bibitem[{Lemaire {et~al.}(1996)Lemaire, Field, \& Maillard}]{lemaire_high}
Lemaire, J.~L., Field, D., \& Maillard, J.~P. 1996, 6

\bibitem[{Lepp \& Dalgarno(1988)}]{lep88}
Lepp, S. \& Dalgarno, A. 1988, \apj, 335, 769

\bibitem[{Malloci {et~al.}(2007)Malloci, Joblin, \& Mulas}]{malloci_-line_2007}
Malloci, G., Joblin, C., \& Mulas, G. 2007, Chemical Physics, 332, 353

\bibitem[{{Malloci} {et~al.}(2005){Malloci}, {Mulas}, {Cappellini},
  {Fiorentini}, \& {Porceddu}}]{Malloci2005}
{Malloci}, G., {Mulas}, G., {Cappellini}, G., {Fiorentini}, V., \& {Porceddu},
  I. 2005, \aap, 432, 585

\bibitem[{{Malloci} {et~al.}(2004){Malloci}, {Mulas}, \& {Joblin}}]{malloci04}
{Malloci}, G., {Mulas}, G., \& {Joblin}, C. 2004, \aap, 426, 105

\bibitem[{{Mathis} {et~al.}(1983){Mathis}, {Mezger}, \& {Panagia}}]{Mathis1983}
{Mathis}, J.~S., {Mezger}, P.~G., \& {Panagia}, N. 1983, \aap, 500, 259

\bibitem[{McKinney {et~al.}(2021)McKinney, Armus, Diaz-Santos, Charmandaris,
  Inami, Song, \& Evans}]{mck21}
McKinney, J., Armus, L., Diaz-Santos, T., {et~al.} 2021, arXiv preprint
  arXiv:2101.01182

\bibitem[{McKinney {et~al.}(2020)McKinney, Pope, Armus, Chary,
  D{\'\i}az-Santos, Dickinson, \& Kirkpatrick}]{mck20}
McKinney, J., Pope, A., Armus, L., {et~al.} 2020, \apj, 892, 119

\bibitem[{Meeus {et~al.}(2012)Meeus, Montesinos, Mendigut{\'\i}a, Kamp, Thi,
  Eiroa, Grady, Mathews, Sandell, Martin-Za{\"\i}di, {et~al.}}]{mee12}
Meeus, G., Montesinos, B., Mendigut{\'\i}a, I., {et~al.} 2012, \aap, 544, A78

\bibitem[{Megeath {et~al.}(2015)Megeath, Gutermuth, Muzerolle, Kryukova, Hora,
  Allen, Flaherty, Hartmann, Myers, Pipher, Stauffer, Young, \& Fazio}]{meg15}
Megeath, S.~T., Gutermuth, R., Muzerolle, J., {et~al.} 2015, 151, 5

\bibitem[{Montillaud {et~al.}(2013)Montillaud, Joblin, \&
  Toublanc}]{montillaud_evolution_2013}
Montillaud, J., Joblin, C., \& Toublanc, D. 2013, \aap, 552, A15

\bibitem[{{Mulas} {et~al.}(2006){Mulas}, {Malloci}, {Joblin}, \&
  {Toublanc}}]{mulas06b}
{Mulas}, G., {Malloci}, G., {Joblin}, C., \& {Toublanc}, D. 2006, \aap, 456,
  161

\bibitem[{Okada {et~al.}(2013)Okada, Pilleri, Berné, Ossenkopf, Fuente,
  Goicoechea, Joblin, Kramer, Röllig, Teyssier, \& van~der
  Tak}]{okada_probing_2013}
Okada, Y., Pilleri, P., Berné, O., {et~al.} 2013, \aap, 553, A2

\bibitem[{Pabst {et~al.}(2021)Pabst, Goicoechea, Hacar, Teyssier, Bern{\'e},
  Wolfire, Higgins, Chambers, Kabanovic, G{\"u}sten, {et~al.}}]{pab21}
Pabst, C., Goicoechea, J., Hacar, A., {et~al.} 2021, arXiv preprint
  arXiv:2111.12363

\bibitem[{Padovani {et~al.}(2009)Padovani, Galli, \&
  Glassgold}]{padovani2009cosmic}
Padovani, M., Galli, D., \& Glassgold, A.~E. 2009, \aap, 501, 619

\bibitem[{Palacios {et~al.}(2010)Palacios, Gebran, Josselin, Martins, Plez,
  Belmas, \& Lebre}]{pal10}
Palacios, A., Gebran, M., Josselin, E., {et~al.} 2010, \aap, 516, A13

\bibitem[{Pilleri {et~al.}(2012)Pilleri, Montillaud, Berné, \&
  Joblin}]{pilleri_evaporating_2012}
Pilleri, P., Montillaud, J., Berné, O., \& Joblin, C. 2012, \aap, 542, A69

\bibitem[{{Racine}(1968)}]{Racine_HD200775}
{Racine}, R. 1968, \aj, 73, 233

\bibitem[{{Rapacioli} {et~al.}(2006){Rapacioli}, {Calvo}, {Joblin}, {Parneix},
  {Toublanc}, \& {Spiegelman}}]{Rapacioli2006}
{Rapacioli}, M., {Calvo}, F., {Joblin}, C., {et~al.} 2006, \aap, 460, 519

\bibitem[{Rapacioli {et~al.}(2005)Rapacioli, Joblin, \&
  Boissel}]{rapacioli_spectroscopy_2005}
Rapacioli, M., Joblin, C., \& Boissel, P. 2005, \aap, 429, 193

\bibitem[{Ricca {et~al.}(2012)Ricca, Bauschlicher, Boersma, Tielens, \&
  Allamandola}]{Ricca_2012}
Ricca, A., Bauschlicher, C.~W., Boersma, C., Tielens, A. G. G.~M., \&
  Allamandola, L.~J. 2012, The Astrophysical Journal, 754, 75

\bibitem[{Rubin {et~al.}(2009)Rubin, Hony, Madden, Tielens, Meixner,
  Indebetouw, Reach, Ginsburg, Kim, Mochizuki, {et~al.}}]{rub09}
Rubin, D., Hony, S., Madden, S., {et~al.} 2009, \aap, 494, 647

\bibitem[{Salas {et~al.}(2019)Salas, Oonk, Emig, Pabst, Toribio,
  R{\"o}ttgering, \& Tielens}]{sal19}
Salas, P., Oonk, J., Emig, K., {et~al.} 2019, \aap, 626, A70

\bibitem[{Salgado {et~al.}(2016)Salgado, Berné, Adams, Herter, Keller, \&
  Tielens}]{salgado_orion_map_eps_2016}
Salgado, F., Berné, O., Adams, J.~D., {et~al.} 2016, \apj, 830, 118

\bibitem[{{Smith} {et~al.}(2007{\natexlab{a}}){Smith}, {Armus}, {Dale},
  {Roussel}, {Sheth}, {Buckalew}, {Jarrett}, {Helou}, \&
  {Kennicutt}}]{Smith_2007_CUBISM}
{Smith}, J.~D.~T., {Armus}, L., {Dale}, D.~A., {et~al.} 2007{\natexlab{a}},
  \pasp, 119, 1133

\bibitem[{{Smith} {et~al.}(2007{\natexlab{b}}){Smith}, {Draine}, {Dale},
  {Moustakas}, {Kennicutt}, {Helou}, {Armus}, {Roussel}, {Sheth}, {Bendo},
  {Buckalew}, {Calzetti}, {Engelbracht}, {Gordon}, {Hollenbach}, {Li},
  {Malhotra}, {Murphy}, \& {Walter}}]{SmithPahfit}
{Smith}, J.~D.~T., {Draine}, B.~T., {Dale}, D.~A., {et~al.} 2007{\natexlab{b}},
  \apj, 656, 770

\bibitem[{{Spaans} {et~al.}(1994){Spaans}, {Tielens}, {van Dishoeck}, \&
  {Bakes}}]{spa94}
{Spaans}, M., {Tielens}, A.~G.~G.~M., {van Dishoeck}, E.~F., \& {Bakes},
  E.~L.~O. 1994, \apj, 437, 270

\bibitem[{Spitzer(2004)}]{spitzer_physical_2004}
Spitzer, L. 2004, Physical processes in the interstellar medium, Physics
  textbook (Wiley)

\bibitem[{Spitzer~Jr(1948)}]{spit48}
Spitzer~Jr, L. 1948, \apj, 107, 6

\bibitem[{Terenin \& Vilessov(1964)}]{ter64}
Terenin, A. \& Vilessov, F. 1964, Advances in Photochemistry, 2, 385

\bibitem[{Tielens(2005)}]{Tielens_book05}
Tielens, A. 2005, The Physics and Chemistry of the Interstellar Medium

\bibitem[{Tielens(2021)}]{tie21}
Tielens, A.~G. 2021, Molecular astrophysics (Cambridge University Press)

\bibitem[{Tielens(2008)}]{tielens_interstellar_2008}
Tielens, A. G. G.~M. 2008, Annual Review of \aap, 46, 289

\bibitem[{Tschurl \& Boesl(2006)}]{Tschur06}
Tschurl, M. \& Boesl, U. 2006, International Journal of Mass Spectrometry,
  249-250, 364, chava Lifshitz Memorial Issue

\bibitem[{{Verstraete} {et~al.}(1990){Verstraete}, {Leger}, {D'Hendecourt},
  {Defourneau}, \& {Dutuit}}]{Verstraete1990}
{Verstraete}, L., {Leger}, A., {D'Hendecourt}, L., {Defourneau}, D., \&
  {Dutuit}, O. 1990, \aap, 237, 436

\bibitem[{Visser {et~al.}(2007)Visser, Geers, Dullemond, Augereau, Pontoppidan,
  \& van Dishoeck}]{vis07}
Visser, R., Geers, V., Dullemond, C., {et~al.} 2007, \aap, 466, 229

\bibitem[{Wahab {et~al.}(2022)Wahab, Pfuderer, Paenurk, \&
  Gershoni-Poranne}]{Wahab22}
Wahab, A., Pfuderer, L., Paenurk, E., \& Gershoni-Poranne, R. 2022, Journal of
  Chemical Information and Modeling, 62, 3704, pMID: 35881922

\bibitem[{{Weingartner} \& {Draine}(2001)}]{weingartner_Draine_01_sizedistrib}
{Weingartner}, J.~C. \& {Draine}, B.~T. 2001, \apj, 548, 296

\bibitem[{Weingartner \& Draine(2001)}]{weingartner_photoelectric_2001}
Weingartner, J.~C. \& Draine, B.~T. 2001, \apjs, 134, 263

\bibitem[{{Wenzel} {et~al.}(2020){Wenzel}, {Joblin}, {Giuliani}, {Rodriguez
  Castillo}, {Mulas}, {Ji}, {Sabbah}, {Quiroga}, {Pe{\~n}a}, \&
  {Nahon}}]{wenzel2020}
{Wenzel}, G., {Joblin}, C., {Giuliani}, A., {et~al.} 2020, \aap, 641, A98

\bibitem[{{Werner} {et~al.}(2004){Werner}, {Uchida}, {Sellgren}, {Marengo},
  {Gordon}, {Morris}, {Houck}, \& {Stansberry}}]{Werner04_NGC7023}
{Werner}, M.~W., {Uchida}, K.~I., {Sellgren}, K., {et~al.} 2004, \apjs, 154,
  309

\bibitem[{Zhen {et~al.}(2016)Zhen, Castillo, Joblin, Mulas, Sabbah, Giuliani,
  Nahon, Martin, Champeaux, \& Mayer}]{zhen_vuv_2016}
Zhen, J., Castillo, S.~R., Joblin, C., {et~al.} 2016, \apj, 822, 113

\end{thebibliography}

\begin{appendix}

\section{Derivation of $\gamma$ and $R_{\rm e}$ in NGC 7023}

\subsection{Derivation of $\gamma$ in the NGC~7023 NW PDR} 
\label{app_derivation_gamma_NGC7023}


In the following, we describe how we derived values for the physical conditions, that is $G_0$, $T$ and $n_{\rm e}$,
necessary to determine $\gamma$ in NGC 7023, for 8 angular distances ranging from 20 to 60'' from the star. 
To derive the radiation field intensity, we used the procedure described in \citet{pilleri_evaporating_2012} and adopted in 
\citet{Joblin2018}. Briefly, this consists in using {  the sum of} two identical synthetic spectra with effective temperatures 
 1.5$\times10^4$\,K from the \citet{Kurucz1993} library and a radius of 10 $R_{\sun}$ for the HD\,200775 
 stars \citep{alecian2008characterization}. 
 An additional extinction of $A_V=1.5$ was applied to 
 account for the presence of dust in the immediate surrounding of the stars. The radiation field intensity 
 as a function of distance, in units of the Habing field, is derived assuming that the star-to-PDR inclination is of 63 degrees,
 which is necessary to reconcile the sky projected distance to the modeled star-to-PDR distance of 0.143\,pc 
 \citep{Joblin2018}. For angular distances from the star above 45'', the effect of dust absorption in the PDR 
 is taken into account using a factor $e^{-\tau(\lambda)}$ where $\tau(\lambda)=\sigma_{ext}(\lambda)N_{\rm H}$,
 with $N_{\rm H}$ the total hydrogen column density along the axis between the PDR and the star.  With a constant density in the PDR of $n_{\rm H}=2\times10^4$\,cm$^{-3}$ \citep{pilleri_evaporating_2012}, the estimated values for  $N_{\rm H}$ 
at positions situated 50'', 55'' et 60'' from the star are  $10^{21}$, $2\times10^{21}$, and $3.2\times10^{21}$\,cm$^{-2}$,
respectively. This corresponds to extinction values of $A_V=0.5, 1$, and $1.6$\,mag, respectively.
The resulting radiation field intensities are computed by integrating the specific intensities in the 
5.17 to 13.6 eV range, and are reported in Table~\ref{tab_parametresphysiques}.
The gas temperature values for the positions up to 45'' from HD 200775 are taken from the study 
of \citet{boulais_stage} who modeled the spatial profile of the [\ion{C}{ii}] 158~$\mu$m line emission using an LTE radiative transfer model {to determine} a stable temperature of $\sim 750$\,K. The values adopted inside 
the PDR, that is above 45'', are from the models of \citet{montillaud_evolution_2013}.
Gas densities $n_{\rm H}$ in the cavity region up to 25'' from the star 
are from \citet{Berne2015}. The authors provided a value of $n_{\rm H}\sim1.5\times10^4$\,cm$^{-3}$ at $\sim 25$'', which we assume to be constant up to the PDR dissociation {front from which} we use a value of $n_{\rm H}\sim2\times10^4$\,cm$^{-3}$ \citep{pilleri_evaporating_2012}. The {gas temperature and density} for all studied positions are reported in 
Table~\ref{tab_parametresphysiques}. The electron density is derived from the gas density by 
considering that all electrons originate from the ionization of carbon, that is $n_{\rm e}=1.6\times10^{-4}n_{\rm H}$\,cm$^{-3}$.
The gas temperature and the electron density can then be used to derive the ionization parameter $\gamma=\frac{G_0 \sqrt{T}}{n_{\rm e}}$ (see values in Table~\ref{tab_parametresphysiques}). \citet{Andrews2016} used $\gamma\sim2\times10^4$\,K$^{1/2}$cm$^3$ at the border of the PDR, ($\sim42''$ from HD\,200775), which is consistent with the values derived in the present study.



\subsection{Derivation of $R_{\rm i}$ in the NGC 7023 NW PDR}
\label{app_derivation_fion_obs_ngc7023}


We limit the study to the field of view (FoV) presented in Fig.~\ref{fig_NGC7023}, which excludes 
the parts of the nebula that are closer than 20'' from the star, and farther than 60''.
The former have to be excluded because other features attributed 
to species such as C$_{60}$ and C$_{60}^+$ \citep{Berne2013}, and a strong thermal continuum 
due to hot dust grains \citep{berne_analysis_2007} are present in the spectra.
This contamination makes a clean analysis of the PAH emission with PAHTAT difficult. 
Similarly, at distances above 60'', the signal-to-noise ratio in the spectra becomes too low
to use PAHTAT. In addition, regions too far from the star 
become optically thick in the mid IR (see Fig.~4 in \citealt{pilleri_evaporating_2012}),
which is also a challenge when using PAHTAT. 
We used the dataset obtained by \citet{Werner04_NGC7023} 
with the IRS instrument aboard \textit{Spitzer}. The data {were previously analyzed by} \citet{berne_analysis_2007} and \citet{boersma_properties_2013}, notably.  
The cube was obtained at a spatial resolution of 
3.6'' with a sampling of pixels of 1.8'' and a spectral resolution of $\sim$ 80, over a spectral range 
ranging from 5.5 to 15 $\mu$m. {  We applied the PAHTAT tool \citep{pilleri_evaporating_2012},
which adjusts spectral templates of PAHs (notably cationic and neutral PAHs) to observations, 
to all spectra in the mid-IR cube to derive the integrated intensities of the PAH$^+$ and PAH$^0$ components
(respectively $I_{\rm{PAH}^+}$ and $I_{\rm{PAH}^0}$) as well as {their sum}
$I_{\rm PAH}=I_{\rm{PAH}^+}+I_{\rm{PAH}^0}$.}
The $R_{\rm i}$ map is then obtained using $I_{\rm{PAH}^+}$ and $I_{\rm{PAH}^0}$ following Eq.~\ref{eq_r_i}.

\subsection{Derivation of $R_{\rm e}$ in the NGC 7023 NW PDR}
\label{app_derivation_gastopah_NGC7023}

\begin{figure*}[!ht]
    \centering
    \includegraphics[width=\hsize]{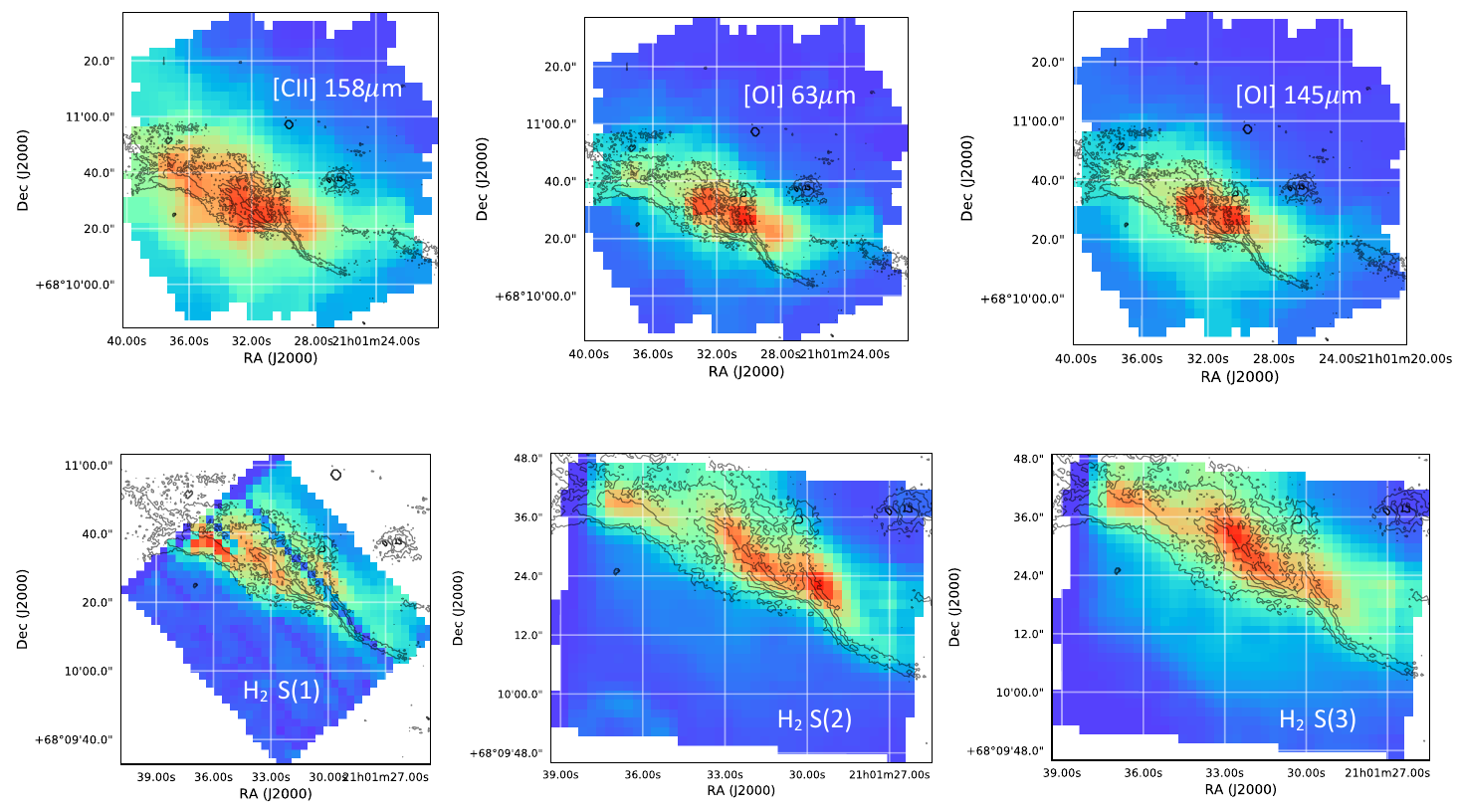}
    \caption{
    Emission maps of {gas cooling.}  Maps {on the upper row} show the 
    [\ion{C}{ii}] line at 158\,$\mu$m, as well as the [\ion{O}{i}] lines at 145\,$\mu$m  and 63\,$\mu$m, which were
    observed with \textit{Herschel}-PACS at a resolution of 11'', 8.8'', and 4.5'' respectively \citep{bernard-salas_spatial_NGC7023_2015}. {Maps on the lower row} show the emission from the H$_2$ 
    rotational S(1), S(2) et S(3) lines at 17.035, 12.278, and 9.7\,$\mu$m, respectively. 
    These maps were extracted from the \textit{Spitzer} archival data using CUBISM \citet{Smith_2007_CUBISM}. 
    The S(2) and S(3) maps are from the SL module of IRS (resolution of 3.6''), 
    and the S(1) map is from the IRS LL module (resolution of 10''). Emission from the H$_2$ v=1-0 S(1) line at 2.121\,$\mu$m is presented in contours as a spatial reference for the dissociation front, but it is not included in the cooling {budget}, since this line is pumped by UV photons and hence does not cool the gas.}
    \label{fig_cooling-maps}
\end{figure*}

Gas cooling in the neutral ISM of galaxies is dominated by emission in fine-structure lines of [\ion{C}{ii}] 
at 158\,$\mu$m, and of [\ion{O}{i}] at 63\,$\mu$m and 145\,$\mu$m. In addition, emission in the rotational 
lines of H$_2$, in the mid-IR, and emission in  CO lines and other molecular species in the far IR
can contribute.  \citet{bernard-salas_spatial_NGC7023_2015} provided a detailed study of gas cooling in the 
far IR in NGC 7023 with \textit{Herschel}-PACS and SPIRE. 
We use their data for the [\ion{C}{ii}] line at 158 $\mu$m, and the [\ion{O}{i}] lines at 63 and 145$\mu$m
(Fig.~\ref{fig_cooling-maps}).
We complement them with the 0-0 S(1), S(2), and S(3) rotational lines of H$_2$ at 17.0, 12.3, and 9.7\,$\mu$m
derived from {\it Spitzer} observations (Fig.~\ref{fig_cooling-maps}). 
The emission maps for the S(2) and S(3) H$_2$ lines were derived 
using PAHTAT \citep{pilleri_evaporating_2012} on the \textit{Spitzer}-IRS SL {data} cube of the NGC~7023 NW PDR obtained by \citet{Werner04_NGC7023}. The map of the S(1) line integrated intensity was extracted using the CUBISM 
software \citep{Smith_2007_CUBISM} on the  LL data cube obtained by \citet{Werner04_NGC7023}.
Overall, the gas line emission is largely dominated by the [\ion{O}{i}] and [\ion{C}{ii}] lines. 
H$_2$ {emission} is important only at the dissociation front (up to 30$\%$ maximum {of the total line emission}), but overall it represents
less than 10\% of the gas cooling. In addition, cooling by H$_2$ emission occurs in regions where gas heating 
is not dominated by PE heating, but rather by the formation of H$_2$, and the collisional de-excitation
of vibrationally pumped H$_2$. Hence, we do not include H$_2$ emission in the gas cooling budget, 
$I_{\rm gas}$ being then the sum of [\ion{O}{i}] and [\ion{C}{ii}] lines emission. 
As for the derivation of the map of $R_{\rm i}$, we limit the study to the FoV 
presented in Fig.~\ref{fig_NGC7023}, which excludes the parts of the nebula that are closer 
than 20'' from the star, and farther than 60''. The $R_{\rm e}$ map is then obtained
from $I_{\rm PAH}$ (Sect.~\ref{app_derivation_fion_obs_ngc7023}) and $I_{\rm gas}$ following Eq.~\ref{eq_r_e}.

\section{Derivation of $R_{\rm e}$ and $\gamma$ in a sample of sources.}
\label{app_derivation_gamma_all_objects}

\subsection{Galactic photodissociation regions}

We include data for the Orion Veil, which is the extended nebular region in front and south of the M42 
region of Orion A, illuminated by the O6 star $\Theta^1$ Ori-C . It is a diffuse nebula where cooling is largely dominated 
by [CII] emission \citep{pab21}. All the data used here were kindly provided by C. Pabst. 
It consists in the table of values of $\gamma$ 
obtained for this region totaling over 11 000 pixels, as well as the values 
of the [CII] line intensities $I_{\rm [CII]}$ derived from SOFIA observations. In addition, we use 
the \textit{Spitzer}-IRAC $8~\mu$m maps of the region provided by \citet{meg15}. 
 Emission in this IRAC filter ($I_{8 \mu \rm m}$) is largely dominated by 
PAH emission, hence, in the absence of spectroscopic measurements, it can be used 
to evaluate $I_{\rm PAH}$. We have used the data of NGC~7023, where $I_{\rm PAH}$ 
is obtained from spectroscopic analysis (Sect.~\ref{app_derivation_fion_obs_ngc7023}) as well as the 
$I_{8 \mu \rm m}$ IRAC maps, and found $I_{\rm PAH} \approx 1.15 \times10^{-7} I_{8 \mu \rm m}$, with 
$I_{\rm PAH}$ in W\,m$^{-2}$\,sr$^{-1}$ and $I_{8 \mu \rm m}$ in MJy\,sr$^{-1}$. This is used to 
derive $I_{\rm PAH}$ from the IRAC data, and $R_{\rm e}$ is then 
given by $\frac{I_{\rm [CII]}}{I_{\rm PAH}+I_{\rm [CII]}}$. The values presented in 
Fig.~\ref{fig_All-objects_obs} and summarized in Table~\ref{tab_all-objects}
are obtained by determining the median and 
first and 99$^{\rm th}$ percentiles for $R_{\rm e}$, and 
median and 25$^{\rm th}$ and 75$^{\rm th}$ percentiles for $\gamma$. The used interval for $\gamma$ is smaller than for $R_{\rm e}$ because of the presence of a larger number of outlier values of $\gamma$.

We include data for the Orion Bar (also illuminated by $\Theta^1$ Ori-C) from \citet{bernard-salas_spatial_orion_2012}
for the cooling lines ([OII], [CII]) at position ($\alpha$ = 5:35:19.778, $\delta$ = -5:25:30.65).
We derive a total intensity for the cooling lines (dominated by [OI] at 63\,$\mu$m) of 
$6.6\times 10^{-5}$ W\,m$^{-2}$\,sr$^{-1}$. 
We extract the value of $I_{8 \mu \rm m}$ at this position both from the IRAC data and available 
IRS spectroscopy, which respectively yields $I_{8 \mu \rm m}=5500$ and $I_{8 \mu \rm m}=7850$ MJy\,sr$^{-1}$,
which gives $I_{\rm PAH} = [6.32 - 9.02] \times 10^{-4}$ W\,m$^{-2}$\,sr$^{-1}$. 
$R_{\rm e}$ is found to fall in the range [0.07-0.11]. 
The range of $\gamma$ values is derived from model results 
for the Orion Bar presented in \citet{Joblin2018}, and corresponding to the 
range of A$_V$, for which the calculated [OI] and [CII] emissions are the dominant cooling lines. 
Values of $R_{\rm e}$  and $\gamma$ for the Orion Bar are 
summarized in Table~\ref{tab_all-objects}.

We include the IC 63 nebula, which is illuminated by the B0.5 star $\gamma$-Cas, 
and for this source $I_{\rm PAH}$ is taken from Table~3 in \citet{pilleri_evaporating_2012}. The values for the cooling lines are taken from 
\citet{and18}. The error on absolute intensities of the cooling lines 
is $\sim$ 30$\%$, including uncertainty on beam filling factor and instrumental error. 
The values of $\gamma$ are from \citet{pab21}.

Ced 201 is a nebula formed by a runaway B9.5 star penetrating a molecular 
cloud in Cepheus. For this source, $I_{\rm PAH}$ is also taken from \citet{pilleri_evaporating_2012}. 
The values for the cooling lines intensities are directly taken from the 
observations available in the Herschel Science Archive and extracted on the central spaxel 
of the PACS instrument. The same error as for IC 63 on the intensity, that is 30\%, is considered. 
The values of $\gamma$ are from \citet{pab21}.

The Horsehead is illuminated by the $\sigma$-Ori AB 
stars of spectral type O9.5V and B0.5V.
The values for $I_{\rm PAH}$ are also taken from \citet{pilleri_evaporating_2012}, and the values for the cooling lines
are also from the observations available in the Herschel Science Archive and
extracted in the central spaxel of PACS (error = 30\%). {To estimate the} values of $\gamma$
we use a {FUV} radiation field of $G_0=100$, a temperature of $T=100-500$\,K,
and {a gas} density of $n_{\rm H}=1-20 \times 10^4$\,cm$^{-3}$ \citep{hab05}. 
This yields $\gamma$ values in the range [30-1400].

\subsection{The diffuse interstellar medium}

For the diffuse interstellar medium, we use the IR emission intensity  for PAHs deduced
by \citet{com11}, that is $I_{\rm PAH}=1.45\times10^{-31} {\rm W}\,{\rm H}^{-1}$
and the cooling measured for the [CII] line from \citet{bou96}, that is $\Lambda_{\rm gas} = 3 \pm0.4 \times10^{-33}{\rm W}\,{\rm H}^{-1}$, which yields  $R_{\rm e}=0.021 \pm 0.002$. We neglect the cooling from [OI] which is marginal for this 
low density medium. The average gas density ranges between 50 and 60 cm$^{-3}$ \citep{bou99}, and the gas temperature is $T\sim~80$~K. The radiation field in the diffuse ISM is $G_0\sim 1.7$. With $n_{\rm e}= 1.6\times 10^{-4} n_{\rm H}$, this yields $\gamma \in [1580-1900]$.

\subsection{The Large Magellanic Cloud}

For the Large Magellanic Cloud (LMC), we rely on the work of 
\citet{leb12} who provide a value of $R_{\rm e}=0.07$,
and quote a factor of 2 uncertainty. We thus adopt $R_{\rm e} \in [0.05-0.01]$.
From Fig.~10 of \citet{rub09}, {we derive} $G_0/n_{\rm e} \in [630-1260]$, which for 
a temperature of 75 K \citep{rub09} yields $\gamma\textbf{} \in [5460 - 10~900]$.

\subsection{Star-forming galaxies}

For star-forming galaxies in the local Universe, we use the ratio of 
the [CII] {emission to the} integrated flux in the 5 to 10  $\mu$m range
provided in \citet{hel01}. Following \citet{mck21}, we apply 
a factor 0.4 to this ratio to obtain $R_{\rm e}$, to account for the fact that PAH emission {is responsible} 
for only about 40$\%$ of the mid-IR emission of galaxies \citep{hel01}.
The range of values of $\gamma$ for the galaxies are taken from \citet{mck21}
using the same approach as for ULIRGs, {which is} based on the relationship of 
\citet{dia17} and presented in the following section.

\subsection{Ultra luminous infrared galaxies}

\citet{mck21} present measurements
of the ratio between the PAH to [CII] emissions, which 
we adopt {to derive} $R_{\rm e}$. 
These authors also provide the values of $G_0/n_{\rm H}$ in 
a large sample of ultra luminous infrared galaxies (ULIRGs).
Using these data, kindly provided in electronic format 
by the authors, we derive the median values of $R_{\rm e}$
and $\gamma$ and 25$^{\rm th}$ and $75^{\rm}$ confidence intervals 
(see values in Table~\ref{tab_all-objects}). We compute $\gamma$
from $G_0/n_{\rm H}$ considering a temperature of 100\,K, in agreement with estimated 
values for $T$ in these galaxies \citep{dia17}, and using 
$n_{\rm e}=1.6\times10^{-4}~n_{\rm H}$.

\subsection{High-redshift galaxy GS IRS20}

\citet{mck20} provide a measurement of the ratio of the [CII] cooling 
line to the intensity of the PAH 6.2 $\mu$m band in the GS IRS20 
galaxy at a redshift $z\approx 1.9$. This value and associated error bar can 
be converted to $R_{\rm e}$ using the calibration factor of 
0.11 obtained by \citet{SmithPahfit} on a large sample
of nearby galaxies for the ratio of the 6.2 $\mu$m band to 
total PAH emission. Additional uncertainty stems from 
the lack of precise measurement of the 11.2 $\mu$m feature 
which implies a possible overestimation of the total PAH 
emission, of a factor of $\sim 2$ (McKinney priv. com.), which we include to obtain
the maximum value of $R_{\rm e}$. 
The range of values for $\gamma$ is obtained 
by using the observed value of the IR surface density of this galaxy
given in \citet{mck20}, that is $\Sigma_{\rm IR} \approx 5\times10^{11}$
L$_{\sun}$kpc$^{-2}$, converted to a value of $G_0/n_{\rm H}$ using 
the law calibrated on ULIRGs by \citet{dia17}, and finally to a value of 
$\gamma$ using a temperature of 100\,K and $n_{\rm e}=1.6\times10^{-4}~n_{\rm H}$.

\subsection{Protoplanetary Disks}

The disks for which we derived $R_{\rm e}$ are Herbig Ae/Be objects, which show 
prominent PAH emission. {Amongst the disks} observed by ISO and presented in \citet{ack04}, {we selected those} for which there 
are \textit{Herschel} measurements of the gas cooling lines \citep{mee12}.
This represents a total of 10 objects, which are listed in Table~\ref{tab_disks}.
For some objects the [OI] 145\,$\mu$m and [CII] lines were not detected.
However this is not an issue since the [OI] 63\,$\mu$m line is by far the most important 
cooling line in these objects, where the gas is dense and highly irradiated. 
The point presented in Fig.~\ref{fig_All-objects_obs} represents the 
median value of $R_{\rm e}$ from Table~\ref{tab_disks} and the 
25$^{\rm th}$ and 75$^{\rm th}$ percentiles. The range of values for $\gamma$ can only be obtained 
indirectly. We used the {results of the} detailed model for PAHs in Herbig Ae/Be stars 
developed by \citet{vis07}. In particular, these authors show that over 
80$\%$ of the PAH emission arises from regions beyond the inner
100 AU of the disk, and their Fig.~5 provides the values for 
$G_0/n_{\rm e}$, {which range} between $10^3$ and 10$^4$ in this region of the disk. For a gas temperature of $\sim 1000$\,K, this implies values for $\gamma$ 
in the range $[3\times 10^4 - 3 \times 10^6]$. The used temperature value is a rough 
estimate, however this choice is not critical 
since $\gamma$ depends {only} on $\sqrt{T}$.

\begin{table*}[]
\centering
\caption{Intensities of gas cooling lines and PAH emission in Herbig Ae/Be protoplanetary disks.}
\begin{tabular}{cccccc} 
\hline \\
Object Name & [OI] 63\,$\mu$m & [OI] 145\,$\mu$m & [CII] 158\,$\mu$m & PAHs & $R_{\rm e}$ \\
\hline \\
AB AUR    & 851.2   & 44.60& 51.00& 17.9 & 0.0053 \\
HD 97048  & 1592.0  &65.6  &106.8 &55.9  &0.0029 \\
HD 100453 &66.10    &-  &- &7.10  &0.0009 \\
HD 100546 &6043.0   &194.7 &203.8 &92.2  &0.0070 \\
HD 135344 &47.90    &-    &- &1.09  &0.0044 \\
HD 141569 &245.3    &24.90 &11.40 &2.60  &0.0108 \\
HD 139614 &44.50    &- &- &10.4  &0.0004 \\
HD 142527 &53.80    &- &- &2.80  &0.0019 \\
HD 163296 &208.4    &- &- &4.10  &0.0051 \\
HD 169142 &91.50    &- &- &10.16 &0.0009 \\
\hline
\end{tabular}
\tablefoot{Intensities for cooling lines are in $10^{-18}$W\,m$^{-2}$ \citep[from][]{mee12}
and for PAHs in $10^{-14}$ W\,m$^{-2}$ \citep[from][]{ack04}. {$R_{\rm e}$ values are calculated using Eq.~\ref{eq_r_e}.}
\label{tab_disks}}
\end{table*}

\section{Derivation of the ionization potentials, ionization yields and partition 
coefficient.}
\label{app_model_gas_heating_PAHs}

\subsection{Ionization potentials}\label{app_IPs}
{Following  \citet{wenzel2020}, the values of the ionization potentials as a function of the charge Z, for $Z\geq0$, are given by:}
\begin{equation}
    IP^{(Z+1)+}=3.9\text{\,eV}+ \frac{1}{4\pi\epsilon_0}[(Z+\frac{1}{2})\frac{e^2}{a}+(Z+2)\frac{e^2}{a}\frac{0.03\text{\,nm}}{a}]\frac{1\text{C}}{e}\text{\,eV,}
    \label{eq_IPestimate}
\end{equation}

where $\epsilon_0$ is the vacuum permittivity, \textit{e} the elementary charge, and \textit{a} the diameter of the
molecule. The change of variable between \textit{a} and the number of carbon atoms is given by (WD01):
\begin{equation}
    a=\sqrt[3]{\frac{N_{\rm C}}{468}} \text{\,nm}.
    \label{eq_size_to_NC}
\end{equation}
For $Z=-1$, we use $IP=0$ (but see Appendix ~\ref{app_threshold}).  

\subsection{Ionization yields}\label{app_IYields}

Several studies provide measurements of ionization yields for neutral PAHs 
\citep{Jochims1996,Verstraete1990} and cationic PAHs \citep{zhen_vuv_2016,wenzel2020}. 
From these studies, empirical laws have been derived to model the ionization yields. 
For neutral species (Z=0), we use the law derived by \citet{Jochims1996} {(energy values in eV)}:

\begin{equation}
    Y(E)=
    \left\{
    \begin{array}{@{}l@{}c@{}l@{}}
        \begin{array}{@{}l@{}}
        \frac{E-IP^+}{9.2}\\
        1\\
        \end{array} & 
        \text{~~~~for} & 
        \begin{array}{r@{}}
        E \leq \mathrm{IP}^{+}+9.2\; \\ 
        E > \mathrm{IP}^{+}+9.2\; \\
        \end{array}
        \label{eq_yieldjoachims}
    \end{array}
    \right.
\end{equation}

For the second ionization {($Z=1$)}, we use the law provided by \citet{wenzel2020} :

\begin{equation}
    Y^\mathrm{+}(N_\mathrm{C},E)=
    \left\{
    \begin{array}{@{}l@{}c@{}l@{}}
        \begin{array}{@{}l@{}}
        0\\
        \frac{\alpha}{11.3-\mathrm{IP}_{2+}}(E -\mathrm{IP}^{2+})\\
        \alpha\\
        \frac{\beta(N_\mathrm{C})-\alpha}{2.1}\,(E - 12.9) + \alpha\\
        \beta(N_\mathrm{C})
        \end{array} & 
        \text{for} & 
        \begin{array}{r@{}}
        E < \mathrm{IP}^{2+}\; \\ 
        \mathrm{IP}^{2+} \leq E < 11.3\; \\
        11.3 \leq E < 12.9\; \\
        12.9 \leq E < 15.0\; \\
        E \geq 15.0,
        \end{array}
        \label{eq_yieldwenzel}
    \end{array}
    \right.
\end{equation}
where $\alpha = 0.3$ and $\beta$ is given by :
\begin{equation}
    \beta(N_\mathrm{C})=
    \left\{
    \begin{array}{@{}lcl@{}}
        \begin{array}{@{}l@{}}
        0.59 + 8.1 \cdot 10^{-3} N_\mathrm{C}\\ 
        1 
        \end{array} & \text{for} &  \begin{array}{r@{}}
        32 \leq N_\mathrm{C} < 50\; \\ 
        N_\mathrm{C} \geq 50.
        \end{array} 
    \end{array}
    \right.
    \label{eq_yieldwenzel2}
\end{equation}
Finally, for anions (Z=-1), we use $Y=1$ following 
\citet{vis07}.

\subsection{{Threshold energy for electron detachment}}
\label{app_threshold}

In Sect.~\ref{subsec_molecular-parameters}, we have assumed that $Y^{-}(E)=1$ at all energies. This approximation can be justified by the low values of both the electronic affinity and the effective photoabsorption cross section at low energies (see Fig.~\ref{fig_sectionsefficaces}). 
Recently, \citet{Iida22} have shown that electron detachment from the pentacene anion, (C$_{22}$H$_{14}^-$), does not occur at the energy given by the electron affinity but is shifted to higher energies due to competition with radiative cooling.
The authors estimated the survival rate of C$_{22}$H$_{14}^-$ to be 50\% for an internal energy of 2.5\,eV and 0\% above 2.8\,eV.
We calculated the corresponding microcanonical temperature to be 775 and 824\,K for 2.5 and 2.8\,eV, respectively. The calculations use the density of states derived from a modified version of the Beyer-Swinehart algorithm \citep{mulas06b} and the harmonic vibrational frequencies from the \textit{Theoretical PAH Spectral Database}. The electron affinity of cirumcoronene is calculated to be similar to that of pentacene (see \textit{Theoretical PAH Spectral Database}). We therefore assumed that efficient electron detachment would occur for temperatures higher than 800\,K, which correspond to internal energies higher than 6\,eV for C$_{54}$H$_{18}^-$. The calculations were performed with the same method as for C$_{22}$H$_{14}^-$ and using the harmonic frequencies from the \textit{NASA Ames PAH IR Spectroscopic Database} \citep{bauschlicher_nasa_2018,Ricca_2012}.

\begin{figure}[!ht]
    \centering
    \includegraphics[width= \hsize]{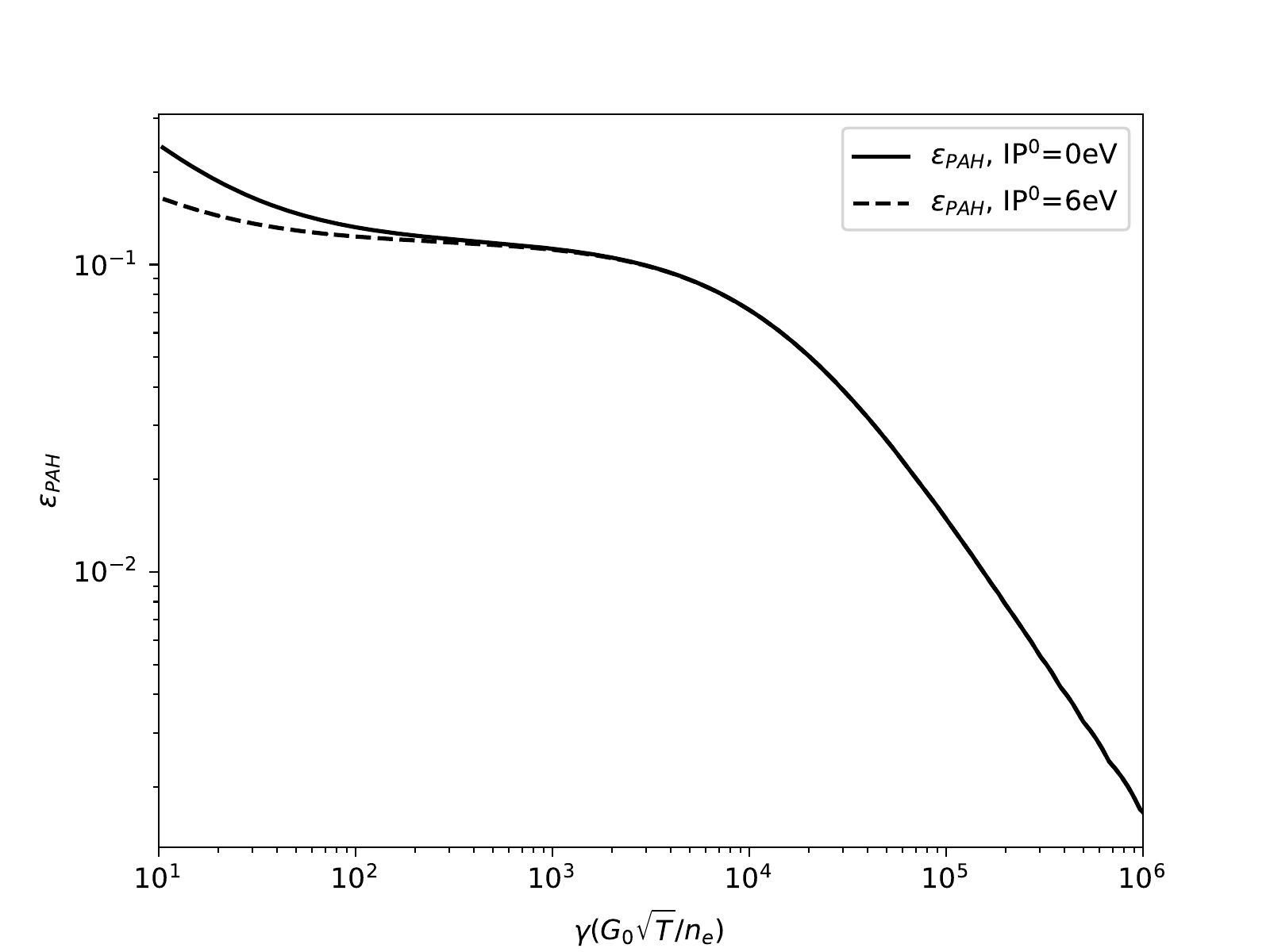}
    \includegraphics[width= \hsize]{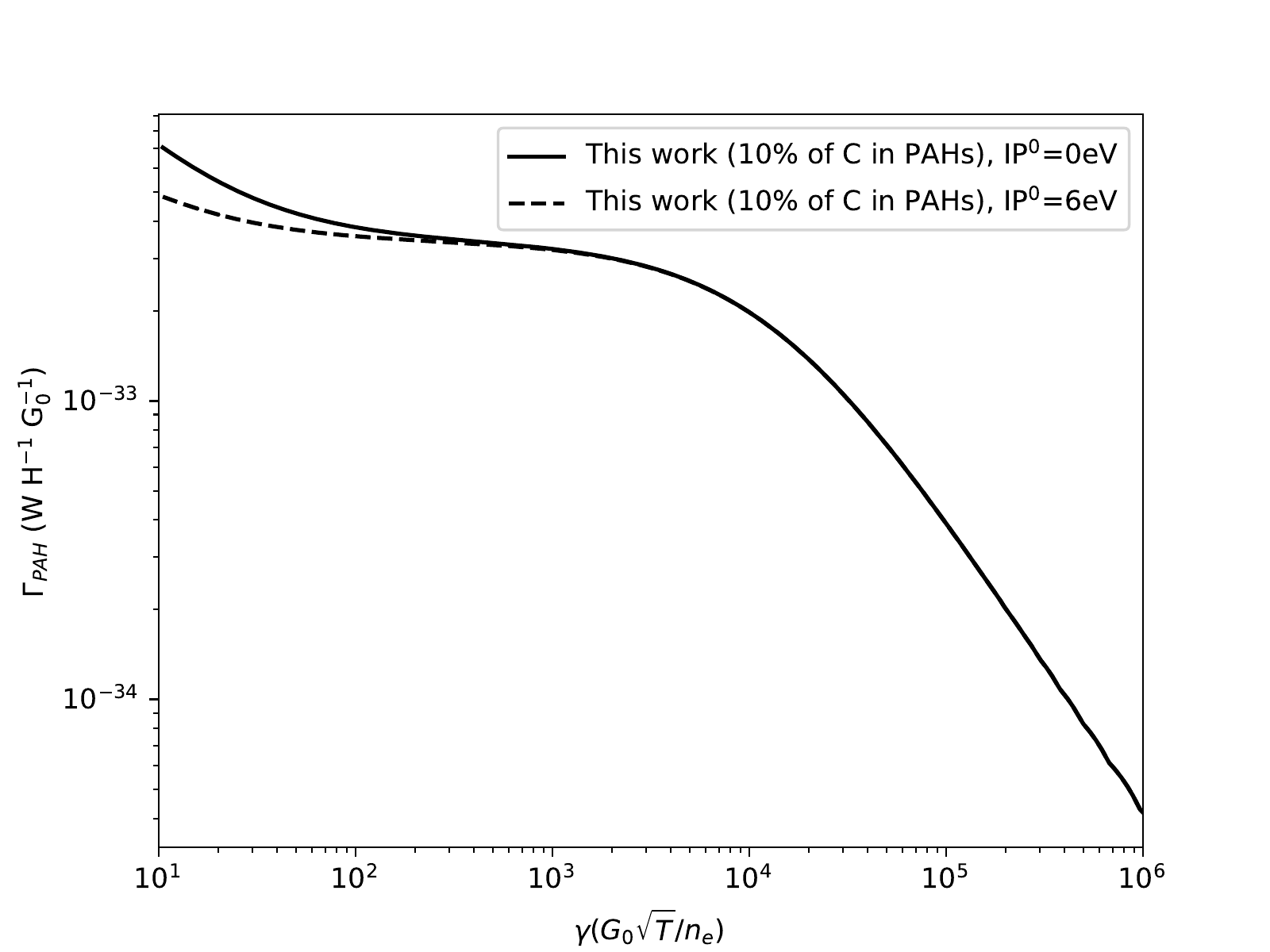}
    \caption{Model derived heating rate $\Gamma_{\rm PAH}$ (upper panel) and 
    PAH photoelectric heating efficiency $\epsilon_{\rm PAH}$ (lower panel) obtained for $IP^0$ = 0 and 6 eV.
    In both cases, the model is computed with $N_C=54$, $T=500$ K, and $T_{\rm Eff} = 3\times 10^4$ K. For the heating rate, we use $f_C=0.1$.}
    \label{fig_threshold}
\end{figure}

In Fig.~\ref{fig_threshold}, we compare the model results for the PAH heating rate $\Gamma_{\rm PAH}$ and the photoelectric heating efficiency of $\epsilon_{\rm PAH}$ as a function of the ionization parameter $\gamma$, which were obtained using $Y^{-}(E)=1$ for $E \ge 6$\,eV (and $Y^{-}(E)=0$ for E < 6\,eV) compared to the case in which $Y^{-}(E)=1$ for all energies. The values of the heating rate and efficiencies are lower by a factor $\sim 1.5$ at $\gamma =10$, by a few percent at $\gamma =100$, and are not affected at $\gamma=100$ and above.
This indicates that the choice of threshold for $Y^{-}(E)$ does not have a significant effect on the results of the model, especially for the large values of $\gamma$ that correspond to the objects studied in this paper (Fig.~\ref{fig_all-objects-epsilon-vs-obs}).

\subsection{{Partition coefficient}}
\label{app_gamma}

\begin{figure}[!ht]
    \centering
    \includegraphics[width= \hsize]{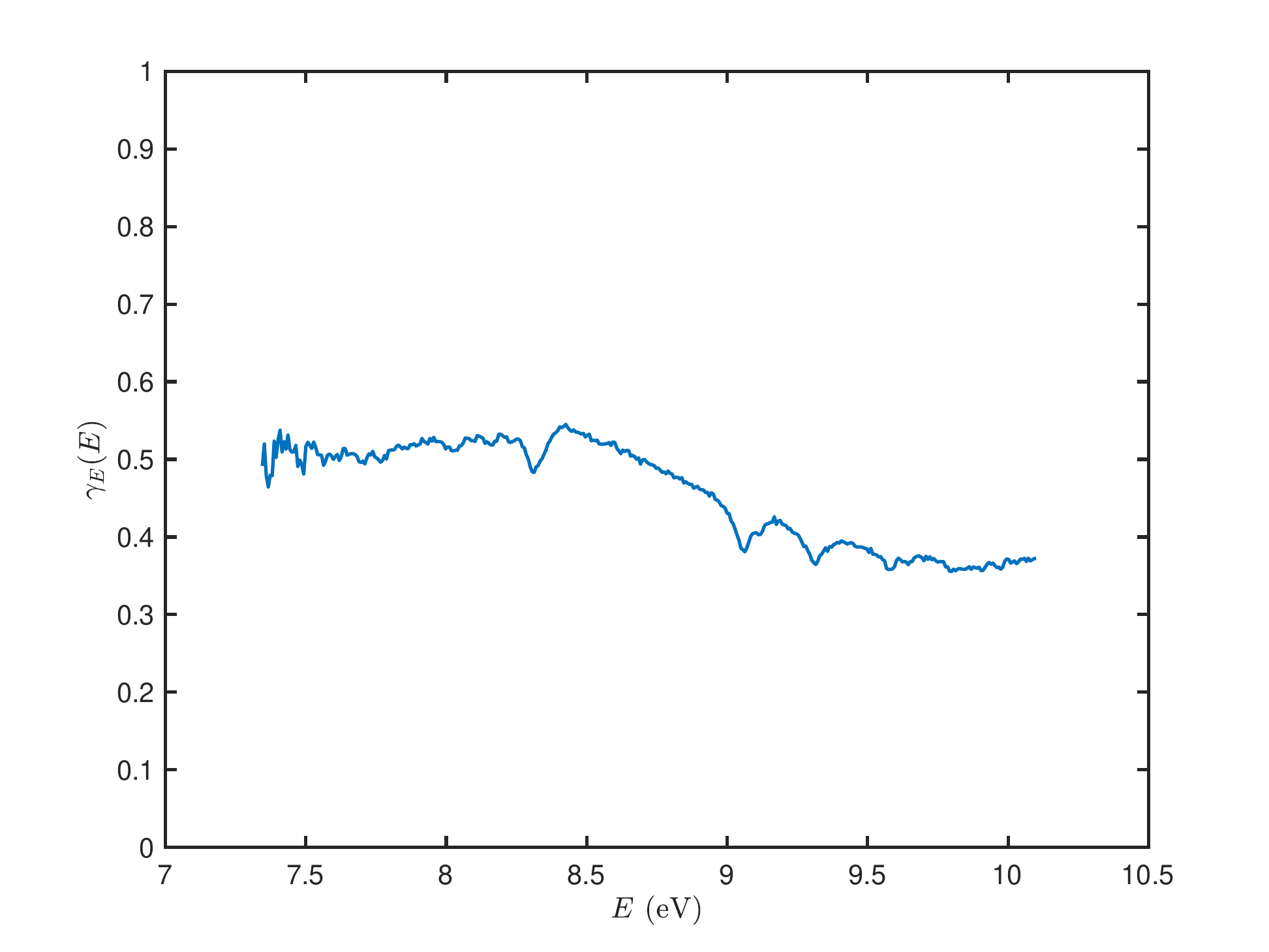}
    \caption{{Fraction of the energy that goes into kinetic energy of the photoelectron}, $\gamma_e(E)$, as a function of incident photon energy, derived from the photoelectron spectroscopic 
    measurements of \citet{brechignac_photoionization_2014} for coronene, C$_{24}$H$_{12}$. The sharp features {are induced by the presence of} auto-ionization states.}
    \label{fig_gamma}
\end{figure}

A value of $\langle \gamma_e (E) \rangle=0.5$ has been adopted in the literature 
\citep{Verstraete1990,Tielens_book05,bakesandtielens94}, based on the 
photoelectron kinetic spectra obtained for specific incident photon 
energies on benzene by \citet{ter64}. We provide a revised 
estimate for this parameter based on the photoelectron spectroscopy 
measurements presented for the coronene molecule in \citet{brechignac_photoionization_2014}. 
In particular, these authors provide (see their Fig.~2), for a given photon energy $E$ {absorbed} in the range 
from 7.3\,eV (IP of coronene) to 10.5\,eV, the probability density function of {kinetic} energy 
of the photoelectrons $E_K(E)$, 
from which $\langle \frac{E_K(E)}{E-IP} \rangle$
can be derived. 
Figure~\ref{fig_gamma} shows the values of $\gamma_e (E) $ for coronene, 
from which $\langle \gamma_e (E) \rangle$  is obtained by averaging over energies, 
yielding $\langle \gamma_e (E) \rangle = 0.46 \pm 0.06~$ (68$\%$ confidence interval), 
in agreement with the classical value of 0.5.

\end{appendix}
\end{document}